\documentclass{aa}

\renewcommand{\linenumbers}{}
\renewcommand{\nolinenumbers}{}
\renewcommand{\runninglinenumbers}{}

\usepackage{graphicx}
\usepackage{txfonts}
\usepackage{xcolor}
\usepackage{booktabs}
\usepackage{orcidlink}
\usepackage{xparse}
\usepackage{enumitem}

\newcommand{\sinopsis}{{\sc sinopsis}\xspace}
\newcommand{\galRPS}{JW100\xspace}
\newcommand{\galCONTROL}{A3376\_B\_0261\xspace}

\begin{document}
\nolinenumbers
 
\title{Distinguishing ram pressure from gravitational interactions: Applying the Size-Shape Difference method to real galaxies}

\author{
  Augusto E. Lassen\inst{1}\thanks{\email{augusto.lassen@inaf.it}}
  \orcidlink{0000-0003-3575-8316}
  \and
  Rory Smith\inst{2,3}\orcidlink{0000-0001-5303-6830}
  \and
  Benedetta Vulcani\inst{1}\orcidlink{0000-0003-0980-149}
  \and
  Stephanie Tonnesen\inst{4}\orcidlink{0000-0002-8710-9206}
  \and
  Paula Calderón-Castillo\inst{2}\orcidlink{0000-0002-7069-113X} 
  \and
  Bianca M. Poggianti\inst{1}\orcidlink{0000-0001-8751-8360}
  \and
  Jacopo Fritz\inst{5}\orcidlink{0000-0002-7042-1965}
  \and
  Koshy George\inst{6}\orcidlink{0000-0002-1734-8455}
  \and
  Alessandro Ignesti\inst{1}\orcidlink{0000-0003-1581-0092}
  \and
  Yara Jaffé\inst{2,3}\orcidlink{0000-0003-2150-1130}
  \and
  Antonino Marasco\inst{1}\orcidlink{0000-0002-5655-6054}
  \and
  Luka Matijevi\'{c}\inst{7,1}\orcidlink{0009-0004-2049-7701}
  \and
  Alessia Moretti\inst{1}\orcidlink{0000-0002-1688-482X}
  \and
  Mario Radovich\inst{1}\orcidlink{0000-0002-3585-866X}
  \and
  Neven Tomi\v{c}i\'{c}\inst{7}\orcidlink{0000-0002-8238-9210}
}

\institute{
  INAF- Osservatorio astronomico di Padova, Vicolo Osservatorio 5, I-35122 Padova, Italy\\
  \email{augusto.lassen@inaf.it}
  \and
  Departamento de Física, Universidad Técnica Federico Santa María, Avenida España 1680, Valparaíso, Chile
  \and
  Millennium Nucleus for Galaxies (MINGAL)
  \and
  Flatiron Institute, Center for Computational Astrophysics, 162 5th Avenue, New York, NY 10010, USA
  \and
  Instituto de Radioastronomía y Astrofísica, UNAM, Campus Morelia, A.P. 3-72, C.P. 58089, Mexico
  \and
  Universit\"ats-Sternwarte\,M\"unchen,\,\,Fakult\"at\,f\"ur\,Physik,\,\,Ludwig-Maximilians-Universit\"at\,M\"unchen,\,\,Scheinerstra{\ss}e\,1,\,81679\,\,M\"unchen, Germany
  \and
  Department of Physics, Faculty of Science, University of Zagreb, Bijeni\v{c}ka Cesta 32, 10000 Zagreb, Croatia
}
   \date{Received June 19, 2025; accepted November 29, 2025}

\titlerunning{SSD method applied to real galaxies}

\abstract
    {In dense environments, mechanisms like ram pressure stripping (RPS) and gravitational interactions can induce the formation of similar morphological features in galaxies, distinguishable only through a detailed study of the stellar properties. While RPS affects recently formed stars through the displacement of the gas disk from which they are formed, gravitational interactions perturb stars of all ages rather similarly.}
   {We present the first observational test of the Size-Shape Difference (SSD) measure, a novel approach, originally designed and validated for simulated galaxies, that quantifies morphological differences between young and intermediate-age stellar populations to distinguish RPS from gravitationally interacting galaxies.}
   {We analyze 67 galaxies from the GASP survey using spatially resolved star formation history derived from the \sinopsis spectral fitting code. In our fiducial model, we compare stellar populations in two age bins ($t < 20\,$Myr and $20\,\rm{Myr} \leqslant t < 570\,\rm{Myr}$) to calculate SSD values. The sample includes confirmed cases of RPS with different stripping intensities, as well as undisturbed and gravitationally interacting galaxies.}   
   {We find that the extreme cases of RPS show SSD values $\sim$3.5$\times$ higher than undisturbed and gravitationally interacting galaxies ($56^{+24}_{-15}$ as compared to $16^{+6}_{-2}$ and $16^{+6}_{-3}$, respectively), confirming simulation predictions. This enhancement reflects RPS-induced asymmetries: the youngest stars are either compressed along the leading edge or displaced into extended tails of cold gas from which they are formed (or both), while older populations remain undisturbed. In contrast, gravitational interactions perturb all stars uniformly, producing lower SSD values.}
   {SSD robustly distinguishes strong RPS cases, even when different age bins are used. This holds even without correcting for disk inclination, or when single-band imaging are used to trace stellar distributions. This makes SSD a promising tool to select RPS candidates for spectroscopic follow-up in upcoming large-scale surveys.}

   \maketitle

\section{Introduction}
\label{sec:intro}

The environment plays a fundamental role in shaping galactic morphologies and regulating their evolution \citep{Dressler1980}. Compared to those in the field, galaxies residing in clusters or groups more frequently exhibit early-type morphologies
\citep{Dressler1980, Whitmore1993, Vulcani2023}, as well as lower star formation rates \citep[SFR;][]{Kennicutt1983, Gavazzi1998, Vulcani2010, Boselli2016, Perez-Milan2023}, and a deficiency in both atomic \citep{Haynes1984, Kenney1989, Catinella2013, Boselli2023} and molecular \citep{Kenney1989, Fumagalli2009} gas content in the spiral population. 

Among the various physical mechanisms galaxies within dense environments may experience\,--\,e.g. tidal interactions, starvation/strangulation and harassment\,--\,ram pressure stripping (RPS) is perhaps the most impactful example, directly altering the cold gas content of the host galaxy. RPS is a hydrodynamical effect resulting from the external pressure exerted by the surrounding hot ($\approx$$10^{7}$\,--\,$10^{8}$\,K) and relatively dense ($\approx$$10^{-3}\,\mathrm{cm}^{-3}$) intracluster medium (ICM) on the interstellar medium (ISM) of the host galaxy \citep{Gunn1972, Poggianti2017, Boselli2022}. Although RPS might enhance star formation during the peak of the stripping phase \citep{Vulcani2018b}, the long-term consequence is the quenching of star formation due to the removal of the cold gas reservoir \citep{Vollmer2001, Vulcani2020a, Cortese2021}.

The pressure exerted by the ICM directly onto the gas disk can truncate it along the leading edge, while additionally creating extended tails on the trailing edge. Due to the spectacular appearance and extent of these ionized gas tails during the peak of the stripping phase, such galaxies are also commonly referred to as jellyfish galaxies, as the tails resemble the tentacles of the sea creature.
As the gas disk is reshaped, the regions in which new stars form also shift. As a consequence, RPS can alter 
both the size and shape of the disk traced by younger stellar populations, but generally has a negligible effect on the stellar populations formed earlier
\citep{Kapferer2009, Bekki2009, Smith2010}\footnote{However, see \citet{Smith2012}, which shows that RPS can exert a drag force on the truncated gas disk, which is subsequently transmitted to the stellar disk and the surrounding central dark matter, displacing them by a few kpc in the direction of the wind.}. In contrast, perturbations to the disk driven by tidal interactions affect the gas and stellar components of a galaxy more equally \citep{Gnedin2003, Boselli2006}. 
As a result, the distribution of the different stellar populations in the galaxy remains much more similar compared to the RPS case.
Moreover, gravitational perturbations tend to redistribute stars over larger areas, lowering the surface brightness of any diffuse morphological signatures that may result.

The striking impact of RPS on the evolution of galaxies within groups and clusters, and the growing interest in deepening our understanding of this process, have motivated concerted observational efforts to gather larger samples of jellyfish galaxies over the past decades \citep{McPartland2016, Poggianti2016, Roberts2021a, Roberts2021b}.
Integral field unit (IFU) spectroscopy consists of a powerful tool for studying the impact of the environment on galaxy evolution, as it enables spatially resolved measurements of the physical conditions and kinematics of both the stellar and gas components of galaxies \citep[][among many others]{Merluzzi2013, Fumagalli2014, Fossati2016, Sanchez2020, Pedrini2022}. 
For instance, hydrodynamical effects such as RPS act upon the gas disk of the galaxy while typically leaving the stellar disk unperturbed. Therefore, IFU data allow for the comparison of both kinematic profiles, serving as an excellent analysis tool to identify disturbances on the gas often caused by RPS in cluster galaxies \citep{Bellhouse2017, Gullieuszik2017, Fritz2017}.
Additionally, IFU observations allow for the spatially resolved modeling of the stellar continuum, which can be used both to separate the gas and stellar components and to reconstruct the star formation history (SFH) of the host galaxy.
This capability is particularly valuable to identify the stellar age distribution within the galaxy \citep{Fritz2017, Bellhouse2021}, whose relevance was highlighted in the previous paragraph.

However, obtaining IFU data is observationally expensive, limiting the sample sizes. 
Spectroscopic follow-up is preceded by the selection of RPS candidates, often based on their apparent morphologies. A complicating factor in this selection is that RPS and gravitational interactions may not act separately.
In fact, hydrodynamical cosmological simulations suggest that both mechanisms often act simultaneously \citep{Marasco2016}, and several observational studies find evidence of galaxies either undergoing both processes or of RPS galaxies exhibiting morphological signatures indicating past gravitational perturbations \citep[][among others]{Vollmer2003, Fritz2017, Sorgho2017, Tomicic2018, Boselli2018, Vulcani2021, Serra2024, Watson2024}.
Furthermore, although the presence of unwinding spiral arms has been often associated with tidal interactions \citep{Dobbs2010, Pettitt2016,Pettitt2017}, at least in the field, recent work by \citet{Bellhouse2021} found evidence for unwinding features produced by RPS alone, providing observational support to early theoretical predictions \citep{Schulz2001,Roediger2014,Steinhauser2016}. This is of particular interest for the selection of RPS candidates, as \citet{Vulcani2022} showed that the fraction of jellyfish galaxies among the cluster population of non-interacting blue late-type galaxies increases significantly when unwinding galaxies are taken into account.

With this in mind, \citet[][hereafter S25]{Smith2025} used simulations of galaxies undergoing either gravitational interactions or RPS to explore how the young (defined as $t < 200\,$Myr in S25) and intermediate-age (defined as $200\,\mathrm{Myr} < t < 400\,\mathrm{Myr}$ in S25) stellar populations respond to each mechanism. Based on these simulations, they developed a novel approach that measures the difference in morphology of the young versus intermediate-age stellar populations, called the ``Size-Shape Difference'' (SSD) method. In S25, the authors demonstrated that the SSD parameter successfully distinguishes cases where RPS decisively shapes the disk morphology in simulations. In this work, we apply the SSD parameter to an observational sample of confirmed RPS galaxies to test whether it can similarly distinguish observed galaxies undergoing strong RPS from those undergoing gravitational interactions, or a control comparison sample of galaxies.

This paper is structured as follows. In Section \ref{sec:sample}, we present and describe the galaxy sample analyzed in this work. In Section \ref{sec:method}, we provide details on how the SSD measurement was applied in S25 and the adaptations required for application to real galaxies.
In Section \ref{sec:results}, we present the main results of this study. In Section \ref{sec:testing_method}, we investigate the impact of changing the input parameters on the SSD measurement. In Section \ref{sec:single_band}, we explore the behavior of the SSD parameter if single-band imaging data are used to trace the distribution of different stellar populations. A summary of the main findings, along with our conclusions, is presented in Section \ref{sec:conclusion}.

Throughout this work, we adopt a \citet{Chabrier2003} Initial Mass Function (IMF) with a 0.1\,$M_{\odot}$\,--\,100$M_{\odot}$ stellar mass range, and standard cosmological parameters of $H_0 = 70\,$km\,s$^{-1}$\,Mpc$^{-1}$, $\Omega_{m} = 0.3$ and $\Omega_{\Lambda} = 0.7$.

\section{The sample}
\label{sec:sample}

\subsection{Sample selection}
To test the SSD method on real galaxies, we require an observational sample in which the stellar ages have been derived in a consistent manner, and one for which environmental influences such as tidal interactions and RPS have been identified. Therefore, we analyze a subsample from the GAs Stripping Phenomena in Galaxies (GASP) survey \citep{Poggianti2017, Poggianti2025}. GASP is an ESO Large Program conducted with the Multi-Unit Spectroscopic Explorer \citep[MUSE,][]{Bacon2010} spectrograph, designed to investigate the physical processes that remove gas from galaxies and their impact on star formation and galaxy evolution. The survey observed a total of 114 galaxies at $z = 0.04\,$--$\,0.07$, including 76 galaxies in cluster environments and 38 in groups, filaments, or isolation. Details on observations, data reduction, and analysis can be found in \citet{Poggianti2017}.

Since our primary focus is distinguishing the effects of RPS from gravitational mechanisms, we analyze a subset of 67 galaxies divided into five categories defined below. RPS galaxies are selected from \citet{Poggianti2025}, which classifies the entire GASP cluster sample based on ionized gas and stellar velocity maps. Using visual assessments of gas disturbance, they confirmed cases where RPS is the dominant mechanism, and assigned each galaxy a stripping strength classification (JType), ranging from very weak/weak to strong and extreme cases (jellyfish galaxies). Our primary sample consists of only the most extreme cases of ram pressure stripping  (JType~=~2, totaling 16 galaxies), although we occasionally include less intense cases (JType~=~1, totaling 18 galaxies) and truncated disks (JType~=~3, totaling 4 galaxies), which represent the final stages of ram pressure stripping.

Although GASP was primarily designed to study hydrodynamical mechanisms, the survey also includes a few galaxies experiencing gravitational interactions \citep{Vulcani2017, Vulcani2021, Poggianti2025}, which are incorporated in our analysis. As extensively described in \citet{Vulcani2021}, these galaxies were identified mainly comparing their gas and stellar velocity fields, as a chaotic stellar kinematics is considered a signature of strong tidal effects or mergers \citep[e.g.,][]{Mihos1993, Struck1999}, while hydrodynamical mechanisms leave the stellar component unaltered. In addition, some of these galaxies have a clear companion which clearly induces tidal interaction \citep{Vulcani2021}.
The considered sample consists of 10 galaxies. We note that this category consists of a very heterogeneous sample which includes both tidal interactions and mergers (from early to late stage mergers) and both field and cluster galaxies, so in principle it is not perfectly comparable to that used in S25. However, applying a strict selection only on tidal interaction galaxies would significantly reduce the sample size, making any comparisons meaningless. 

Finally, we include a control sample of 19 galaxies that are not subject to strong environmental effects and can therefore be considered undisturbed. They indeed do not show any clear disturbances in their stellar and ionized gas distributions. These galaxies have JType~=~0 or 0.3 if part of the cluster sample \citep{Poggianti2025}, or are classified as field undisturbed galaxies \citep{Vulcani2018b, Vulcani2020}.
Alternatively, we tested a classification based on both the H$\alpha$ luminosity of the tails ($L(\mathrm{H}\alpha)$) and the fraction of H$\alpha$ emission outside the approximate extent of the stellar disk \citep[$f_{\mathrm{H}\alpha}^{\mathrm{out}}$; see Fig.~2 in][]{Poggianti2025}. Since this alternative classification yields consistent conclusions, we adopt the widely used JType classification scheme described in detail by \citet{Poggianti2025}.

\subsection{Data analysis}
\label{sec:data_analysis}

The spectrophotometric code \sinopsis \citep{Fritz2014, Fritz2017} is used to obtain spatially resolved estimates of stellar masses and the average star formation rate, as well as the total mass formed within four age bins (representing the star formation history, SFH): 
SFR1 (ongoing star formation) $\equiv$  $t < 2 \times 10^7$ yr, SFR2~$\equiv$~2$\times 10^7 \leqslant t[\mathrm{yr}] < 5.7 \times 10^8$, SFR3~$\equiv$~5.7$\times 10^8 \leqslant t[\mathrm{yr}] < 5.7 \times 10^9$, and SFR4 $\equiv$ $t \geqslant 5.7 \times 10^9$ yr.
Although combinations of different age bins are tested, in this study we primarily focus on the two youngest bins, which we refer to as young and intermediate, respectively. The youngest bin is well-suited to trace stars formed from the stripped gas, while the intermediate bin offers a reasonable timescale over which RPS may have begun producing the extended tails observed today \citep{Smith2022a}.
The total integrated stellar masses are then obtained by adding the values of all the spaxels belonging to each galaxy, as described in \citet{Vulcani2018b}.

Structural parameters -- including inclination, position angle, and effective radius -- were derived in \citet{Franchetto2020} using $I$-band images produced by the MUSE pipeline integrating the reduced data cubes with the $I$-band filter response curve. This band samples the reddest part of the spectrum, resulting in a smooth light profile that is less affected by recently formed young stars and the ionized gas emission associated with them. The effective radius ($R_e$) was obtained by analyzing the azimuthally averaged surface brightness profile, while the position angle (PA) and inclination ($i$) were determined from disk isophotes. For a few galaxies where this procedure did not converge, structural parameters were unavailable. In these cases, we estimate $R_e$ using the mass-size relation provided in Appendix A of \citet{Franchetto2020}, which is based on the GASP sample and minimizes systematic uncertainties.

\section{Methodology}
\label{sec:method}

\subsection{The original method}
\label{subsec:ssd_sims}
In this section, we provide details of the simulations presented in S25, describing how the SSD parameter is measured, and expanding on the physical concepts -- briefly introduced in the introduction -- that enable this parameter to distinguish between RPS and gravitational interactions. Throughout the paper, we will use SSD$_{\mathrm{S25}}$ to refer to the methodology as implemented in S25.

In S25, two galaxies with identical stellar disk masses ($M_{\star} = 1\,\times\,10^{10}\,M_{\odot}$) and gas fractions ($f_g = 0.125$) 
were simulated, each subjected to a different environmental effect. In the first case, the fiducial simulated system consists of a slow-speed (150\,km\,s$^{-1}$) encounter with a less massive (1:4 mass ratio) companion devoid of gas. Different encounter speeds were tested, but higher speeds result in interaction timescales that are too short to induce significant morphological perturbations in the host galaxy. The use of a companion devoid of gas ensures that only gravitational effects are at play. In the second fiducial setup, a wind-tunnel approach was used to simulate the effects of RPS. The two galaxies, initialized with the conditions described above, are followed for up to 2.5\,Gyr after the start of the simulation. For details on the sub-grid recipes implemented and simulation setup, we refer the reader to S25.

Although the evolution of stellar particles is tracked continuously over time, S25 adopts the time windows of $t < 200$\,Myr and $200\,\mathrm{Myr} < t < 400\,\mathrm{Myr}$ as fiducial references to compare the spatial distribution of young and intermediate-age stellar populations in both simulated galaxies, respectively.
As stated in Sect.~\ref{sec:intro}, gravitational interactions and RPS are expected to produce distinct age distributions in stellar populations within this time range. In the case of gravitational interactions, all the star particles are affected more similarly, leading to comparable sizes and shapes for the stellar population distribution. RPS, by contrast, is a hydrodynamical effect, thus acting directly on the cold gas distribution of the galaxy, truncating the gas disk along the leading edge while forming extended tails of stripped gas in the trailing edge.
These distinct age distributions are the key physical idea explored by the SSD$_{\mathrm{S25}}$ measure to distinguish the dominant mechanism in action.

To measure the SSD$_{\mathrm{S25}}$, the coordinates of the disk center must first be determined.
In S25, this was done by using the intermediate-age stars to identify the peak of local density and select it as the disk center.
Each image was divided into angular slices about the disk center. The radius containing 50\%, 75\%, 90\%, 95\%, and 99\% of the total number of stellar particles within each angular slice was calculated -- these are called the ``Lagrangian radii''. The SSD$_{\mathrm{S25}}$ parameter measures the differences between the Lagrangian radii of the young stellar particles and those of the intermediate-age stellar particles. In practice, the SSD$_{\mathrm{S25}}$ is a single value, computed as the sum of the absolute differences between the Lagrangian radii:

\begin{equation}
    \mathrm{SSD}_{\mathrm{S25}} = \sum\limits_{\theta=0}^{2\pi}\,\sum\limits_{i} \left | R_{i}^{\mathrm{young}} (\theta) - R_{i}^{\mathrm{int.}} (\theta) \right |
    \label{eq:ssd}
\end{equation}

\noindent where $R_{i}^{\mathrm{young}}$ and $R_{i}^{\mathrm{int.}}$ are the $i$th Lagrangian radii calculated on the image corresponding to the distribution of young and intermediate-age stars, respectively. As mentioned, in S25 they used the radii containing 50\%, 75\%, 90\%, 95\%, and 99\% of the total number of stellar particles within each angular slice.
Therefore, SSD$_{\mathrm{S25}}$ quantifies the differences between the size and shape of stellar populations of different ages. As consequence, a galaxy in which the young and older stars have identical spatial distributions would yield SSD~=~0.

In S25, SSD$_{\mathrm{S25}}$ is computed over 24 angular slices (i.e., $\Delta\theta = 15^\circ$). While the absolute value of SSD may increase with the number of angular slices, we show in Sect.~\ref{subsec:vary_degbins} that normalized (i.e., SSD divided by the number of slices) SSD values are robust against changes in $\Delta\theta$. Although in principle a normalized form could be adopted, we keep the form of Eq.~\ref{eq:ssd}  to remain consistent with the primary goal of this work: testing the SSD$_{\mathrm{S25}}$ method on observed galaxies.

When applied to a simulated disk galaxy, the SSD$_{\mathrm{S25}}$ value was found to remain low (comparable to an isolated control case), in the case of tidal interactions. This is because the morphological response of the young and intermediate-age stellar populations to the tidal interaction was relatively similar in terms of their overall morphological shapes and sizes. In contrast, when the same model galaxy was exposed to RPS, the morphology of the young stars differed significantly from the intermediate-age stars, due to the stripping of the disk gas. This leads to a significantly higher SSD$_{\mathrm{S25}}$ value for the RPS galaxy.

S25 successfully demonstrated the ability of the SSD$_{\mathrm{S25}}$ parameter to identify the presence of RPS as long as it is sufficiently strong to affect star formation within the stellar disk. But, so far, SSD$_{\mathrm{S25}}$ was only proven on a simulated disk galaxy. In the following sections, we illustrate how the SSD$_{\mathrm{S25}}$ methodology can be adapted to be applied to an observational sample of galaxies, namely the GASP sample.

\begin{figure*}
    \centering
    \includegraphics[scale=0.5]{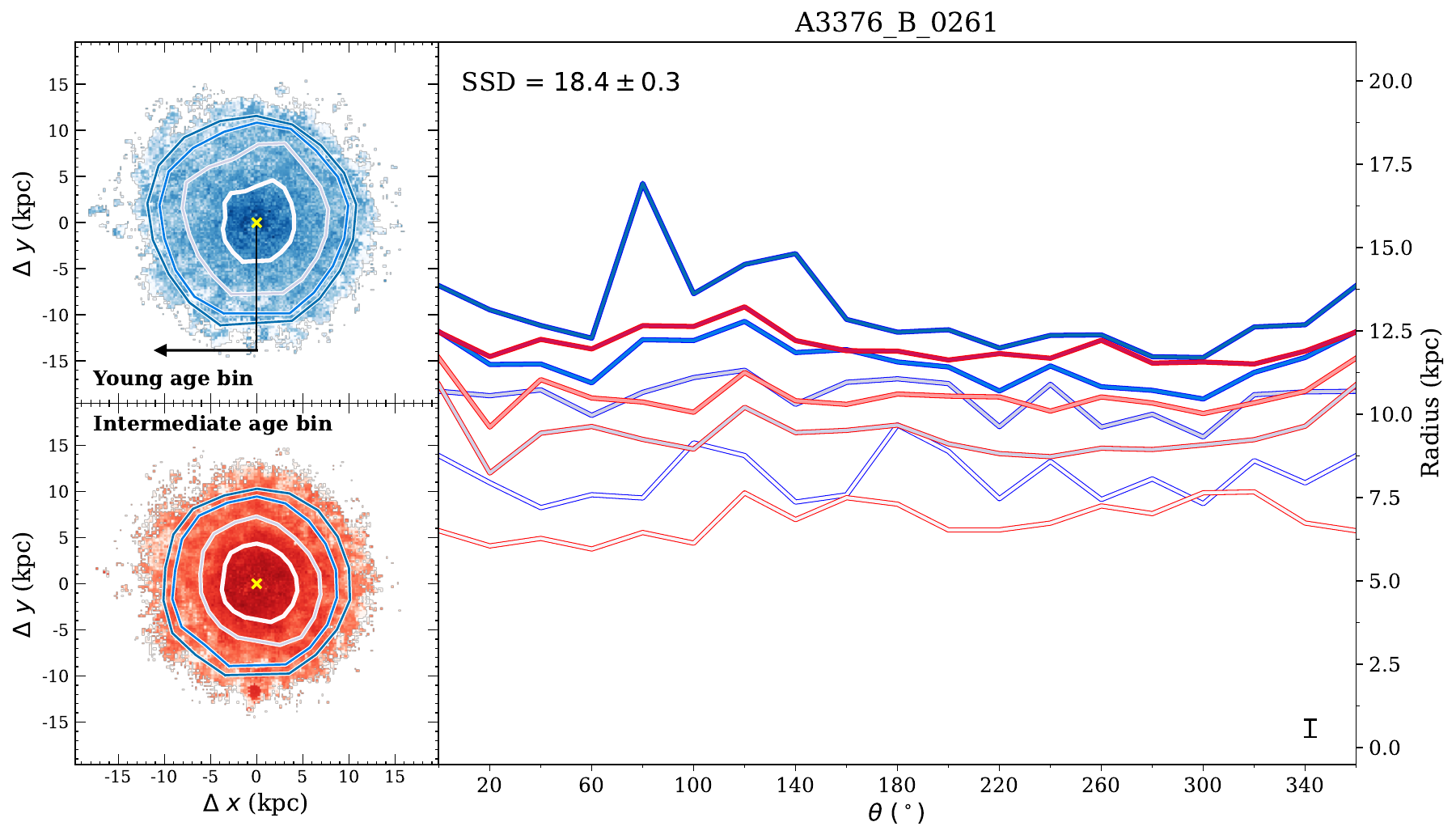}
    \vspace{0.5cm}
    \includegraphics[scale=0.5]{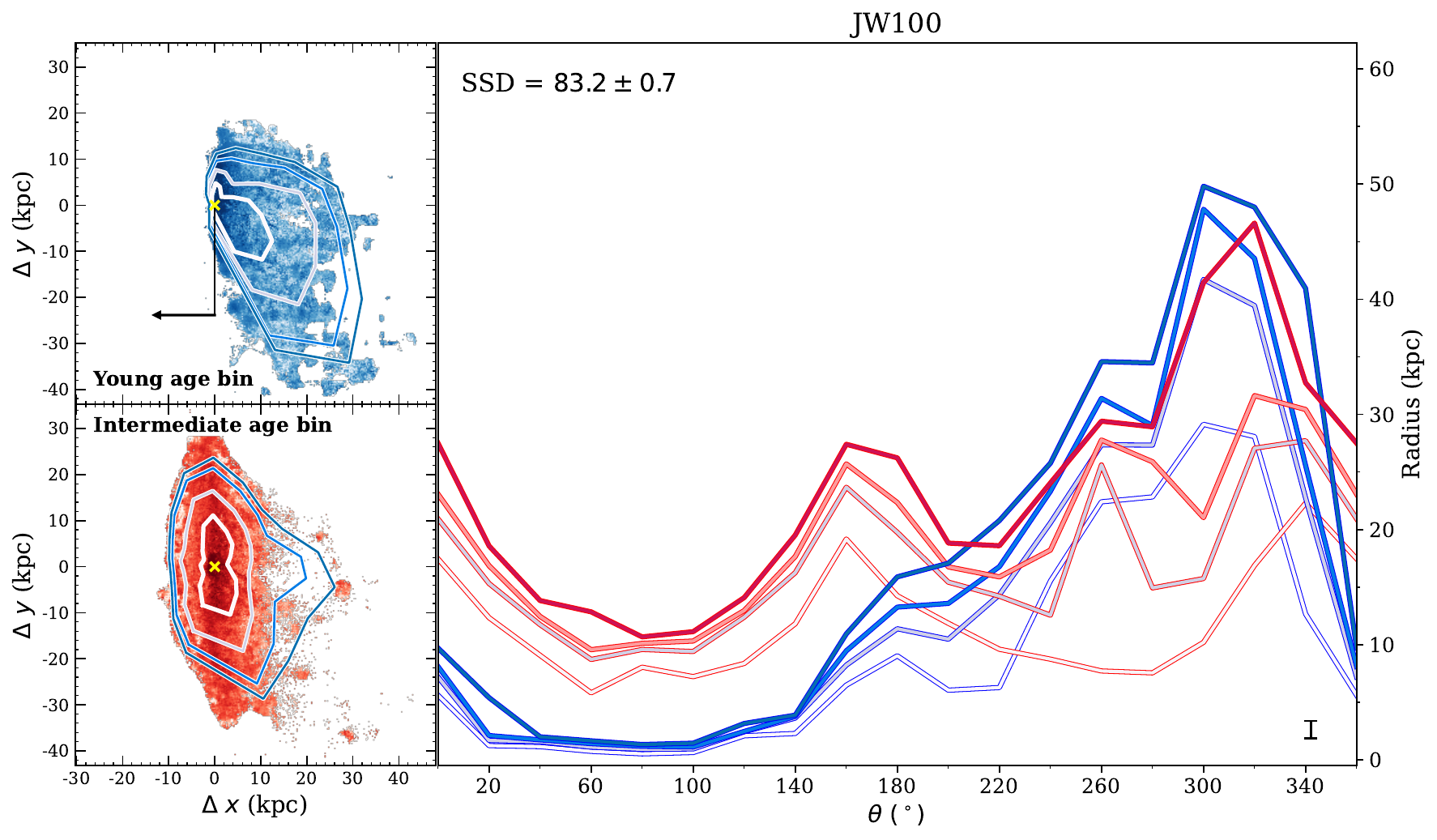}
    \vspace{-0.8cm}
    \caption{SSD method applied to GASP galaxies. \textit{Top:} Spatially resolved maps of star formation rate surface density (SFRD) for the control galaxy with no identified environmental effects, \galCONTROL, obtained with \sinopsis. The top-left frame shows the $t < 20\,$Myr SFRD, while the bottom frame corresponds to the age interval $20\,\mathrm{Myr} \leqslant t < 570\,\mathrm{Myr}$. The color scale is logarithmic with SFRD within the range $-4 \lesssim \log(\mathrm{SFRD}) \,[M_{\odot}\,\mathrm{yr}^{-1}\,\mathrm{kpc}^{-2}] \lesssim -0.5$ . The black line with an arrow indicates the starting point and rotation direction of the varying angle slices (i.e., beginning from South in clockwise rotation).
    Blue and red contours show the Lagrangian radii enclosing 75\%, 90\%, 95\% and 99\% of the total SFRD within each angular slice for the young and intermediate age bins, respectively. The lines grow in a bottom-to-top orientation, following the cumulative order of the Lagrangian radii (i.e, 75\%, 90\%, 95\% and 99\%).
    The black bars on the bottom-right corner of each frame correspond to the mean error of all Lagrangian radii for each case.
    Angle steps of $\Delta \theta = 20^\circ$ were adopted. \textit{Bottom:} Same as top, but for the JType~=~2 galaxy, \galRPS. Top and bottom frames share color scales.
    \label{fig:SSD_GASP}}    
\end{figure*}

\subsection{Applying the SSD to the GASP sample}
\label{subsec:ssd_gasp}
In an observational sample of galaxies, a few adaptations of the original method are required to measure the SSD. For a start, it is no longer possible to count star particles within each angular slice, as it was done in S25 using the simulation output.

Instead, we use the spatially resolved SFR averaged within the age intervals defined in Sect.~\ref{sec:data_analysis}, derived from \sinopsis. These values are converted into star formation rate surface density (SFRD) maps by dividing the SFR in each pixel by its physical area, computed at the luminosity distance of the corresponding galaxy. As the fiducial case, we use the age bins of $t < 20\,$Myr -- that we call ``young'' -- and $20\,\mathrm{Myr} \leqslant t < 570\,\mathrm{Myr}$ -- that we will call ``intermediate'' -- as tracers of the young and intermediate stellar populations, respectively (see Section \ref{sec:data_analysis} for details).
The effect of varying the age bin used to trace the intermediate-age stellar populations is considered in Sect.~\ref{subsec:vary_agebins}.
Alternatively, in Appendix \ref{app:Appendix_Ha} we show the effect of using SFRD derived directly from H$\alpha$ emission line to trace the distribution of the youngest stellar populations. However, like in S25, we find that modifying the age bins does not significantly impact the final SSD values.

For each galaxy, to calculate the Lagrangian radii in the SFRD maps, we use the radii enclosing 75\%, 90\%, 95\% and 99\% of the total SFRD within each angular slice as our fiducial model.
We choose not to include the 50\% Lagrangian radius, as tests showed that this particular Lagrangian radius can be more sensitive to the presence of central features such as AGN and/or bars, making the interpretation of the results harder. Given the point-source nature of AGN emission at our typical spatial resolution, its effect is expected to be distributed nearly uniformly across all angular slices once the innermost Lagrangian radii are excluded. A more detailed discussion of the impact of host galaxy properties on the SSD values is presented in  Sect.~\ref{subsec:galaxy_props}.
The effect of varying the combination of Lagrangian radii used is considered in Sect.~\ref{sec:testing_method}. In fact, calculating the cumulative SFRD inside each slice is nearly equivalent to summing the total masses of stars, given the fixed time bins. As discussed in Sect.~\ref{subsec:vary_degbins}, we adopt angular slices of $20^\circ$, although we show that our results are not very sensitive to this choice.

We emphasize that, in observations, it is essential to account for the presence of foreground and background sources. If these are not masked, they can dominate the flux within a specific angular slice, leading to artificial peaks in the Lagrangian radii that may inflate the final SSD measurement.
Stars were masked before measuring the SFH of the GASP galaxies using \sinopsis. Since \sinopsis requires stellar velocities as input, a Voronoi binning tessellation \citep{Cappellari2003} was applied to all galaxies targeting S/N~=~10 across the range $\lambda_{\mathrm{rest}} \lesssim 7000$\,\AA\, prior to determining the stellar velocities with the  Penalized Pixel-Fitting (\textsc{pPXF}) code \citep{Cappellari2017}. For further details, we refer the reader to \citet{Poggianti2017, Fritz2017}. 
Using the stellar velocities as input, \sinopsis fit is carried out in the individual spaxels. To prevent the inclusion of low S/N spaxels in the calculation of the Lagrangian radii, a (S/N)$_{\mathrm{H}\alpha} \geqslant 3$ threshold is imposed for spaxels belonging to the youngest age bin (SFR1), while for the older age bins (SFR2, SFR3) only spaxels with S/N$\geqslant 3$ over the continuum are considered. The S/N over the continuum is calculated considering the wavelength range of 5210\AA $\leqslant \lambda_{\mathrm{rest}} \leqslant$ 5310\AA\, \citep[defined in][]{Fritz2017}.
Additionally, to eliminate remaining spurious spaxels, we mask out regions smaller than 10 contiguous pixels.

Furthermore, galaxies span a wide range of physical and sky sizes. As a result, larger galaxies will naturally exhibit larger Lagrangian radii and, in turn, may exhibit higher SSD values unless this size effect is accounted for. To establish a SSD value that is less sensitive to galaxy size, enabling a fair comparison across the galaxy sample, we normalize the SSD by each galaxy apparent size on the sky, given by their effective radii ($R_{e}$, in arcsec). This normalization was not necessary in S25, where the size of the simulated galaxy was fixed.

In Fig.~\ref{fig:SSD_GASP}, we illustrate the application of the method to two representative GASP galaxies.
The top and bottom left panels display the SFRD maps derived with \sinopsis for the young and intermediate bins, respectively. The right panels show the 75\%, 90\%, 95\% and 99\% Lagrangian radii (from bottom to top). Blue and red lines represent the young and intermediate components, respectively.
In the bottom-right corner of each right-hand panel in Fig.~\ref{fig:SSD_GASP}, a black bar illustrates the mean uncertainty associated with all Lagrangian radii measured.
These uncertainties are estimated using 2000 random samples of the cumulative SFRD values within each angle slice, drawn from a normal distribution with a spread determined by the pixel-wise SFRD mean error. The SFRD mean error is calculated as the symmetric difference between the pixel-wise upper and lower SFRD boundaries provided by \textsc{sinopsis}, which are derived using an adaptive simulated annealing algorithm. For further details on the derivation of these estimates, we refer the reader to \citet{Fritz2007, Fritz2017}, as well as Section~9.12 of the \textsc{sinopsis} manual\footnote{\label{foot:sinop_manual}\url{https://www.irya.unam.mx/gente/j.fritz/REPOSITORY/SINOPSIS/sinopsis_manual.pdf}}.

In the upper panel of Fig.~\ref{fig:SSD_GASP}, we consider \galCONTROL, which is classified as a control galaxy \citep{Vulcani2018a, Poggianti2025}. It presents a regular morphology, and shows no visible indications that it is undergoing ram pressure stripping. If we consider the right panel, all of the Lagrangian radii of the young (blue) and intermediate (red) age populations are similar, as expected for a galaxy that is not undergoing any external physical mechanisms. 

\begin{table*}[ht!]
\centering
\caption{Properties and SSD value for one representative galaxy of each class. The full version of this table is available at the CDS.}
\label{tab:minitab}
\begin{tabular}{llllll}
\toprule
\toprule
ID & RA (ICRS) & DEC (ICRS) & $z$ & Class & SSD \\
\midrule
    \galCONTROL & 06:00:13.68 & -39:34:49.2 & 0.050596 & Undisturbed & $18.4\pm0.3$ \\
    JO128 & 12:54:56.84 & -29:50:11.2 & 0.049981 & JType 1 & $15.0\pm0.3$ \\
    JO153 &  13:28:15.15 & -31:01:57.9 & 0.046865 & Grav. interactions & $10.1\pm0.3$ \\
    JO23 & 01:08:08.10 & -15:30:41.8 & 0.055097 & JType 3 & $32.3\pm0.9$ \\
    \galRPS & 23:36:25.06 & +21:09:02.5 & 0.061891 & JType 2 & $83.2\pm0.7$ \\
\bottomrule
\end{tabular}

\end{table*}

In contrast, in the lower panel of Fig.~\ref{fig:SSD_GASP}, we consider \galRPS. This is a well-known case of strong RPS, classified as a JType~=~2 galaxy \citep{Poggianti2019a, Poggianti2019b, Poggianti2025}. For $\theta \lesssim 150^\circ$ (the left-hand side of the left panels), we can see that the star-forming disk has been strongly truncated with respect to the intermediate-age population, due to the ram pressure striking the disk at these locations. Similarly, in the right panel, we can clearly see that the blue lines all fall below the red ones, for the same reason. This difference increases the SSD value. Meanwhile, in the south-west side of the galaxy, a tail can be seen in the left panels. Here, the Lagrangian radii corresponding to the younger populations tend to extend beyond their intermediate age counterparts.
Consequently, due to differences both on the leading edge and within the tail, the resulting SSD parameter of \galRPS ($83.2\pm0.7$) is significantly larger than that of \galCONTROL ($18.4\pm0.3$).

\section{Results}
\label{sec:results}

\subsection{Comparing SSD values across galaxy types}
\label{subsec:ssd_comparison}
We are now in a position to measure the SSD values for all GASP galaxies, following the methodology described in Sect.~\ref{subsec:ssd_gasp} and illustrated by Fig.~\ref{fig:SSD_GASP}. This allows us to test whether the approach introduced by S25 also holds for observations, and whether galaxies undergoing different physical mechanisms exhibit systematically distinct SSD values. We stress that the classification into ram pressure stripped, gravitational interactions and undisturbed galaxies is mainly based on the ionized gas and stellar kinematics, hence without considering the galaxy star formation histories. The classification is hence completely independent from the SSD-based criteria presented in the previous section.

\begin{figure*}
    \centering
    \includegraphics[width=\linewidth]{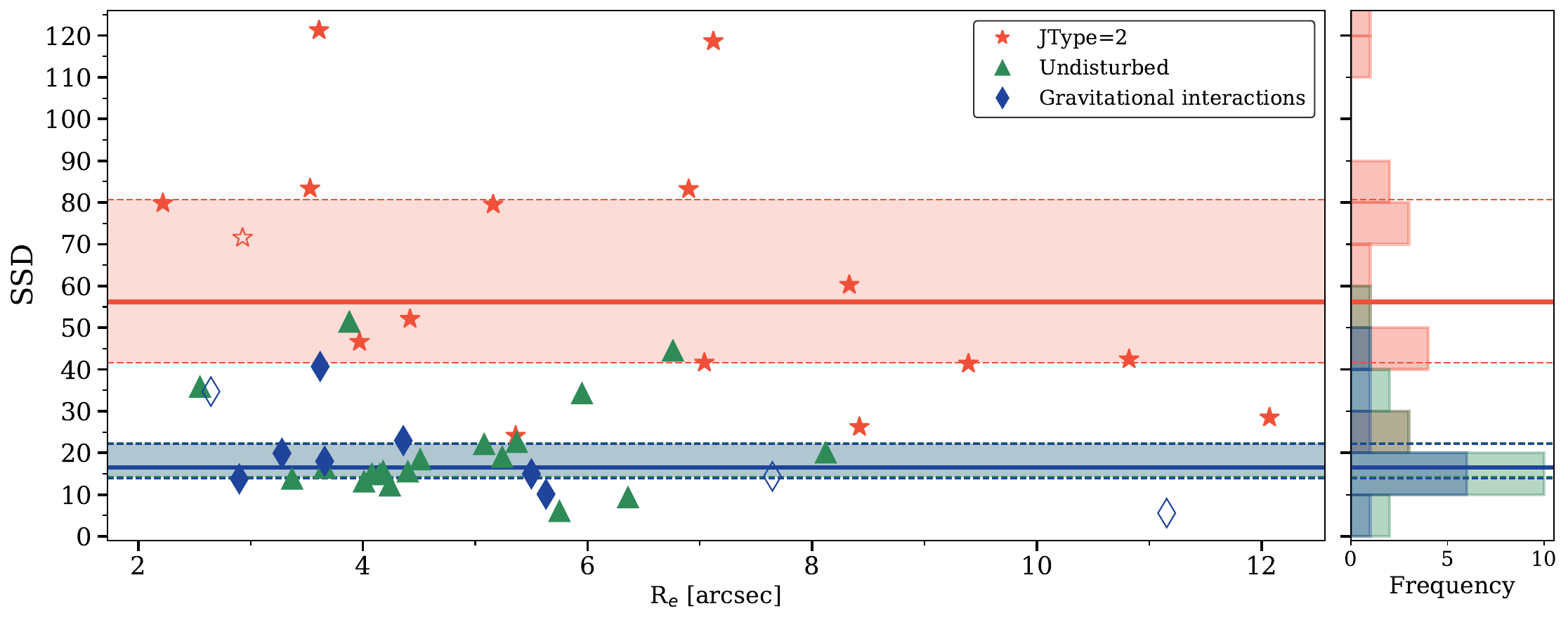}
    \vspace{-0.6cm}
    \caption{Distribution of $R_e$-normalized SSD values as a function of $R_e$ for JType~=~2 (red stars), undisturbed (green triangles), and gravitationally interacting (blue diamonds) galaxies. Empty symbols indicate galaxies for which $R_e$ was estimated using the empirical $R_e$–$M_{\star}$ relation from \citet{Franchetto2020}.
    Solid lines represent the median SSD values for each galaxy type, while the shaded regions correspond to the interquartile range (25\%\,--\,75\%). Dashed lines correspond to the 25\% and 75\% of each galaxy type.
    The SSD value distributions, along with their medians and interquartile ranges, are also shown in the histogram on the right-hand panel.}
    \label{fig:moneyplot}
\end{figure*}

Table~\ref{tab:minitab} presents the SSD values for one representative galaxy from each category defined in Sect.\ref{sec:data_analysis}. Figure~\ref{fig:moneyplot} shows the SSD values as a function of $R_e$ for JType~=~2 (star symbols), undisturbed (triangles), and gravitationally interacting (diamonds) galaxies. These three categories are selected because they are expected to show the most distinct SSD values: undisturbed and gravitationally interacting galaxies are likely to exhibit similarly low SSD values, whereas JType~=~2 galaxies -- representing the strongest cases of RPS -- are expected to show elevated SSD values.
We remind the reader that results obtained for the gravitationally interacting galaxies must be taken with caution as this sample is notably heterogeneous (see Sect.~\ref{sec:sample}). Empty symbols represent galaxies for which $R_e$ is estimated indirectly using the calibration with stellar mass from \citet{Franchetto2020}.

The three solid lines, also shown in the right-hand histogram, represent the median SSD values for each galaxy type -- 56, 16, and 16 for JType~=~2, undisturbed and gravitationally interacting galaxies, respectively -- while the shaded regions indicate the interquartile range (25\%\,--\,75\%) of SSD values within each group. Fig.~\ref{fig:moneyplot} shows that JType~=~2 galaxies exhibit systematically higher SSD values, without overlap between their interquartile range and those of the other two galaxy classes. In contrast, undisturbed and gravitationally interacting galaxies display similarly low and more tightly clustered SSD values, with narrower interquartile ranges than those of the JType~=~2 galaxies, as shown in the right-hand histogram of Fig.~\ref{fig:moneyplot}.
Additionally, as desired, Fig.~\ref{fig:moneyplot} shows no clear dependence between SSD and $R_e$.

\begin{figure}[!ht]
    \centering
    \includegraphics[width=\columnwidth] {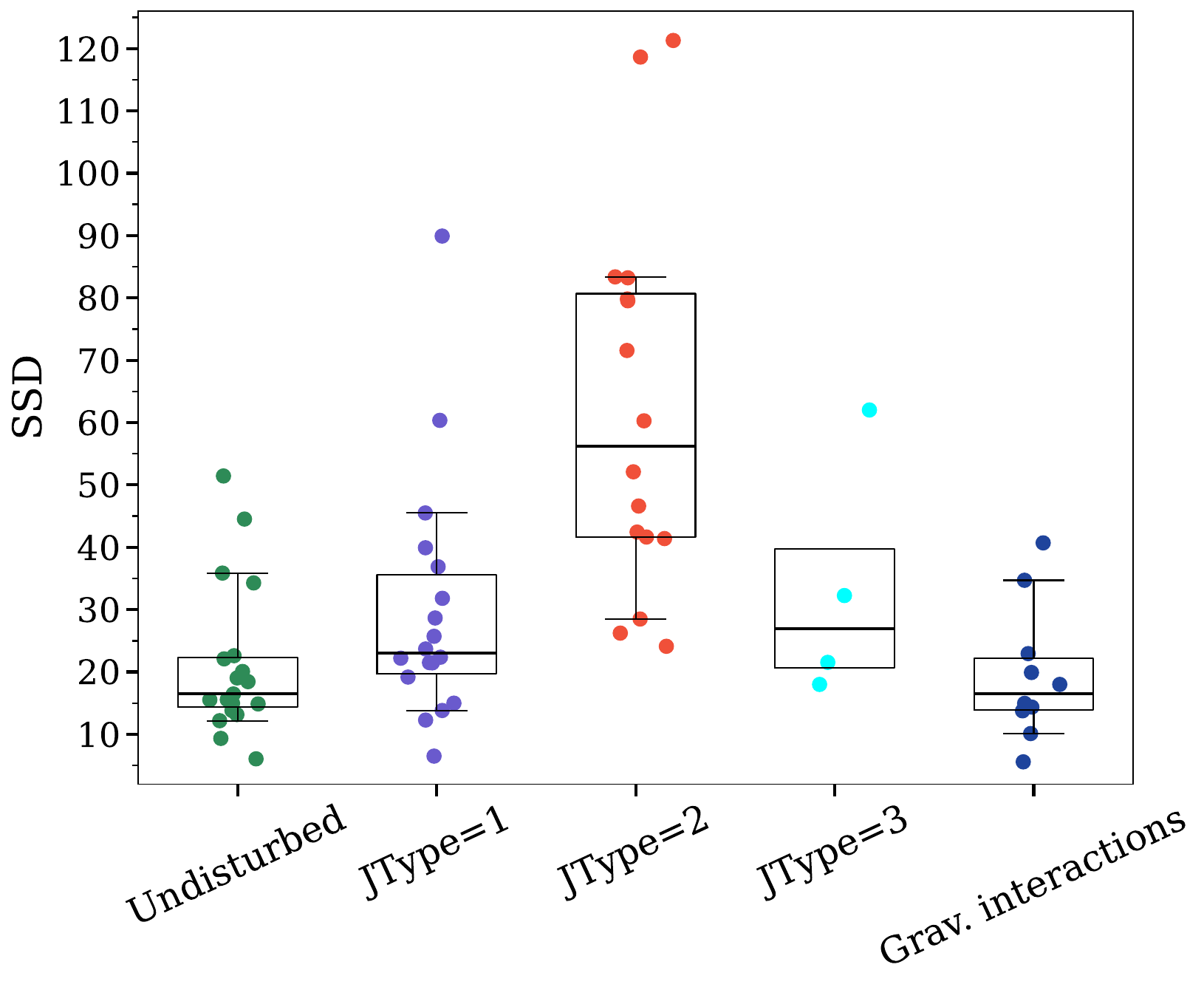}
    \vspace{-0.6cm}
    \caption{Boxplot showing the normalized distribution of SSD values for undisturbed galaxies (green), JType=1 (purple), JType=2 (red), JType=3 (cyan) and gravitationally interacting galaxies (blue). Each point corresponds to the SSD measure of an individual galaxy. A minor shift was applied in the $x$-axis for clarity. The box sizes represent the interquartile range (25\%\,--\,75\%) for each galaxy type, while the whiskers indicate the 10\%\,--\,90\% range of each distribution.}
    \label{fig:cfr_types}
\end{figure}

Aside from the 45 galaxies shown in Fig.~\ref{fig:moneyplot}, it is also important to examine the SSD values of the entire sample analyzed in this work.
This expanded inspection enables comparisons across galaxy types and helps assess whether a clear trend exists between SSD and JType classification. Figure~\ref{fig:cfr_types} presents this comparison using boxplots that show the SSD distribution for all galaxies in the sub-sample analyzed in this work, grouped by galaxy type.
JType~=~1 galaxies, which are objects showing clear signs of ram pressure stripping, but having shorter tails than JType~=~2 galaxies, exhibit  median SSD values slightly higher than undisturbed galaxies. This suggests that the SSD value is sensitive to extended tails and less sensitive for more subtle RPS cases.

In contrast, JType~=~3 galaxies (truncated disks) have SSD values more similar to the  JType~=~2 objects than to the undisturbed sample. In truncated disks, the spatial distribution of the youngest stellar populations should indeed be more concentrated compared to older age bins, hence a high SSD was expected and indeed measured.
We stress that this particular category includes galaxies with relatively high disk inclinations ($i = 80.7^\circ$, $76.3^\circ$, $72.6^\circ$, and $62.3^\circ$), which can directly affect both the predictive power of SSD (inclination effects are discussed in detail in Appendix \ref{app:inclination}) and the derived SFHs, as in highly inclined systems, stellar population properties are averaged along a longer line of sight.
Unlike jellyfish galaxies (e.g., the bottom panel of Fig.~\ref{fig:SSD_GASP}), truncated disks typically exhibit older stellar populations extending farther outward than younger ones across most angular slices, which also contributes to elevated SSD values.
We stress that the definition of $\mathrm{SSD}_{\mathrm{S25}}$ from Eq.~\ref{eq:ssd} is based on the absolute differences between Lagrangian radii in each angle slice, making it sensitive to size-shape differences in the spatial distribution of different stellar populations, regardless of whether younger stars are more centrally concentrated than older ones. While alternative metrics could be explored to attempt distinguishing between ongoing and past RPS events, this work focuses on applying the SSD method — successfully tested on simulated galaxies — to observed systems. Although the sample of truncated disks is small, the SSD method also appears effective in detecting imprints of past RPS episodes on the size-shape difference of stellar populations within the host galaxy.

\subsection{Varying the age bins}
\label{subsec:vary_agebins}
\begin{figure}
    \centering
    \includegraphics[width=\columnwidth]{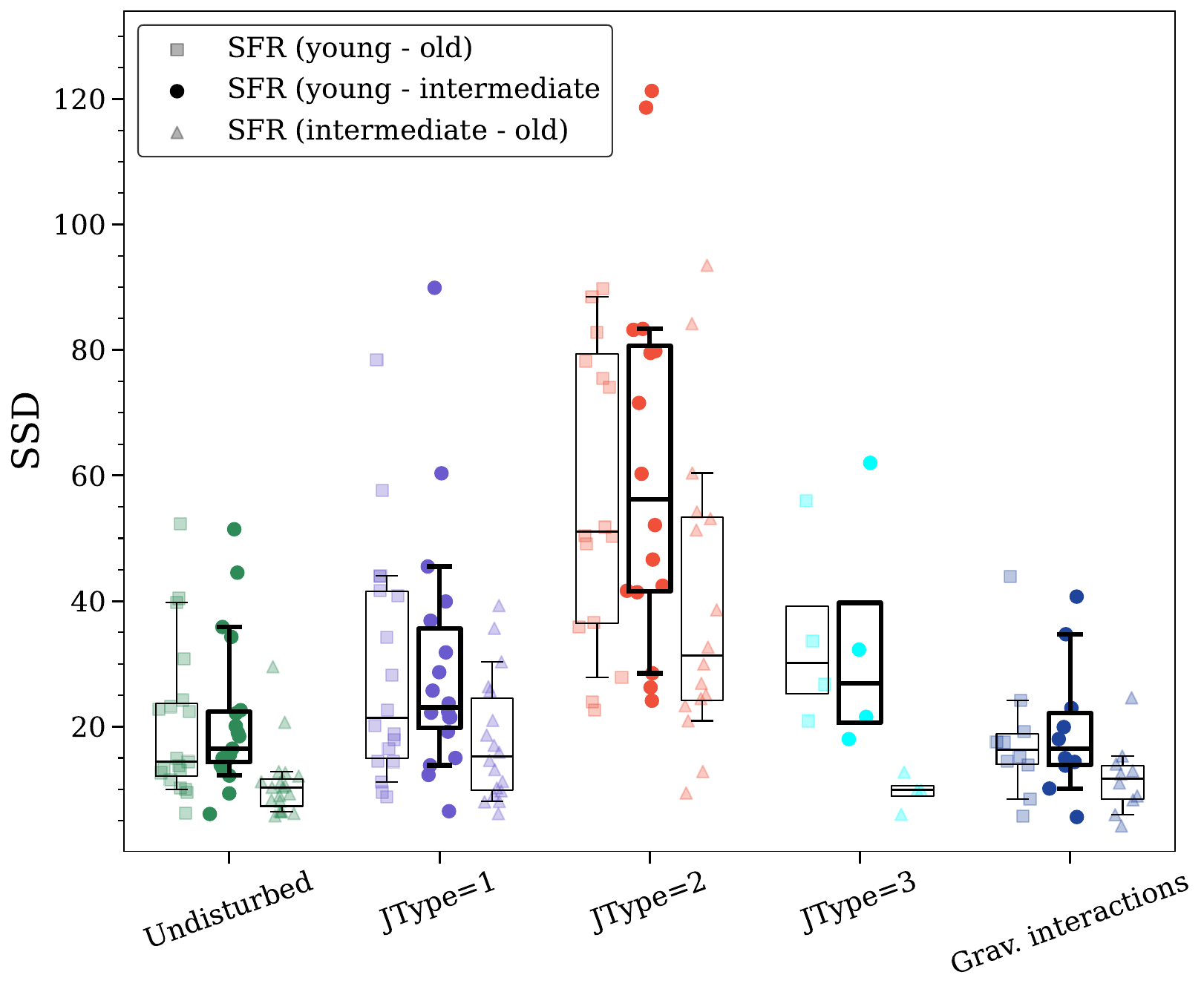}
   \vspace{-0.5cm}
    \caption{SSD values across different galaxy categories, comparing results obtained when different age bins are used to trace the stellar distribution, as indicated in the legend.
    Thicker lines show out fiducial value, shown in Fig.~\ref{fig:cfr_types}. Colors and symbols are as in Fig.~\ref{fig:cfr_types}.}
    \label{fig:cfr_age}
\end{figure}

As stated in Sect.~\ref{subsec:ssd_gasp}, this work compares the distributions of young and intermediate-age stellar populations, which correspond to our fiducial model. To test the impact of different age bins on the SSD results, we compare the fiducial SSD values to two alternative cases in which we use the age bin of $570\,\mathrm{Myr} \leqslant t < 6\,\mathrm{Gyr}$ -- that we call ``old'' (see Section \ref{sec:sample}) -- as the older stellar component, either keeping the young age bin or using the intermediate age bin as the younger stellar component. 
This comparison is shown in Fig.~\ref{fig:cfr_age}. We do not include the $t \gtrsim 6\,$Gyr age bin in the comparison, as in this case galaxy morphology can also be significantly affected by galaxy growth. 

The higher SSD values observed for the JType~=~2 galaxies compared to other galaxy categories suggest the fiducial age bins reflect the distinct distribution of young stars formed from the gas that has been affected by RP, compared to the older stars that formed before the galaxy's interaction with the ICM.
However, the intermediate-age stellar population used in the fiducial model is likely to include stars that formed after the beginning of RPS. Considering that the typical cluster crossing time is on the order of a Gyr, most of the stars formed in RPS galaxies in the last 570\,Myr (the upper age of the intermediate-age bin) would have done so while the galaxy was undergoing RP. Assuming that the extended tail of an infalling galaxy undergoing RPS developed steadily over the course of a few Gyr, one would expect even higher SSD values when an older age bin is used.

In the first case (young vs. old), we find that although individual SSD values may shift relative to the fiducial case, the overall distributions remain remarkably similar, with the difference between JType~=~2 and undisturbed galaxies clearly preserved (no overlap between the interquartile range of the distributions in either case). The fact that Fig.~\ref{fig:cfr_age} shows a very similar average SSD distribution for JType~=~2 galaxies when using either the young vs. old or fiducial age bins -- along with significantly reduced values for the case of intermediate vs. old -- may indicate that most of the star formation within the strong tails likely occurred on timescales shorter than 570\,Myr, even though the tails themselves can survive for longer periods \citep[see, for instance,][]{Bellhouse2019}. This interpretation is supported by previous studies showing that most JType~=~2 galaxies are not only on their first infall, but are already at or near pericentric passage, thus lowering the time they have spent in the densest cluster regions \citep{Jaffe2018, Salinas2024}. Further support comes from idealized simulations by \citet{Tonnesen2012}, which predict that star formation in stripped tails occurs predominantly in high-pressure environments such as cluster centers, a claim supported by observational findings from \citet{Jaffe2018}, who report that JType~=~2 galaxies tend to reside near the centers of massive clusters.

In the second case (intermediate vs. old), the SSD values lie below the fiducial case regardless of the galaxy category. The smallest changes seem to occur among gravitationally interacting galaxies, as expected, given that all stellar populations are expected to be similarly affected by gravitational interactions.
For JType~=~2 galaxies, the median SSD shows a significant (from $\sim$56 to $\sim$31) decrease compared to the other two cases, with the median value below the interquartile range of the fiducial case.
This reduced difference is expected, as SFR3 averages the star formation over $600\,\rm{Myr} \lesssim t \lesssim 6\,\rm{Gyr}$, while the extended stripped tails are thought to have developed only over the past few Gyr \citep{Smith2022a}.

Interestingly, the JType~=~3 galaxies (truncated H$\alpha$ profiles relative to the stellar disk) show a systematic strong drop in SSD values when the old age bin is used (from a median of $\sim$26 in the fiducial case to $\sim$10 in the intermediate vs. old case). 
Although this category contains only a few galaxies, the fact that both size and shape of their stellar distribution are very similar in the intermediate and old age bins suggests that truncation has occurred only in the period probed by the intermediate age bin.
This finding provides interesting insights into post-pericenter evolution, suggesting that star formation in the remaining disk must have ceased within $\sim$500\,Myr after pericenter passage. Otherwise, we would expect to observe JType~=~3 galaxies showing disk truncation between the timescales spanned by the intermediate and old (SFR2 and SFR3) age bins.
As a direct consequence of the quenching, the ancient truncated disks would not have been identified as JType~=~3 galaxies.

Varying the stellar age bins used for the SSD measure allows us to confirm that our results are not strongly dependent on capturing a small window in time, and as long as the ``younger'' population includes stars younger than a few 100\,Myrs, we can find significant differences between RPS galaxies and either undisturbed or those undergoing gravitational interactions. By comparing the SSD measured with the different age bins, we also gain insight into the length of time that these stages would be identified, confirming previous work that long star-forming tails are most likely to be seen nearer to the pericenter of satellite orbits.

We stress that deriving physical properties of stellar populations from integrated stellar emission through the full-spectral fitting technique is inherently subject to degeneracies, such as the well-known age–metallicity degeneracy, for instance. As a consequence, it is difficult to quantify precisely the amount of stellar mass that may be mistakenly attributed to a given age bin. By adopting only four age bins -- thus deliberately decreasing the overall age-resolution power of the method -- \sinopsis aims to minimize this effect, as extensively discussed in \citet{Fritz2007} using integrated spectra from the WINGS survey \citep{Fasano2006}. In the fiducial case (SFR1 vs. SFR2), this issue is significantly mitigated by the self-consistent nebular treatment implemented in \sinopsis, since stellar populations older than $\sim$20\,Myr no longer produce enough UV photons to ionize Hydrogen. This is not the case considering the SFR2 vs. SFR3 case presented in Fig.~\ref{fig:cfr_age}. A quantitative assessment of the level of cross-contamination between these age bins is beyond the scope of this paper and will be addressed in a forthcoming study (Rivas-Sánchez et al., in prep.).

\subsection{Effects of galaxy properties}
\label{subsec:galaxy_props}
In this subsection, we investigate whether other galaxy properties -- such as the presence of a bar or an active galactic nucleus (AGN) -- influence the measured SSD values. We also examine the dependence of SSD on stellar mass, given its correlation with galaxy size.

\begin{figure}
    \centering
    \includegraphics[width=\columnwidth]{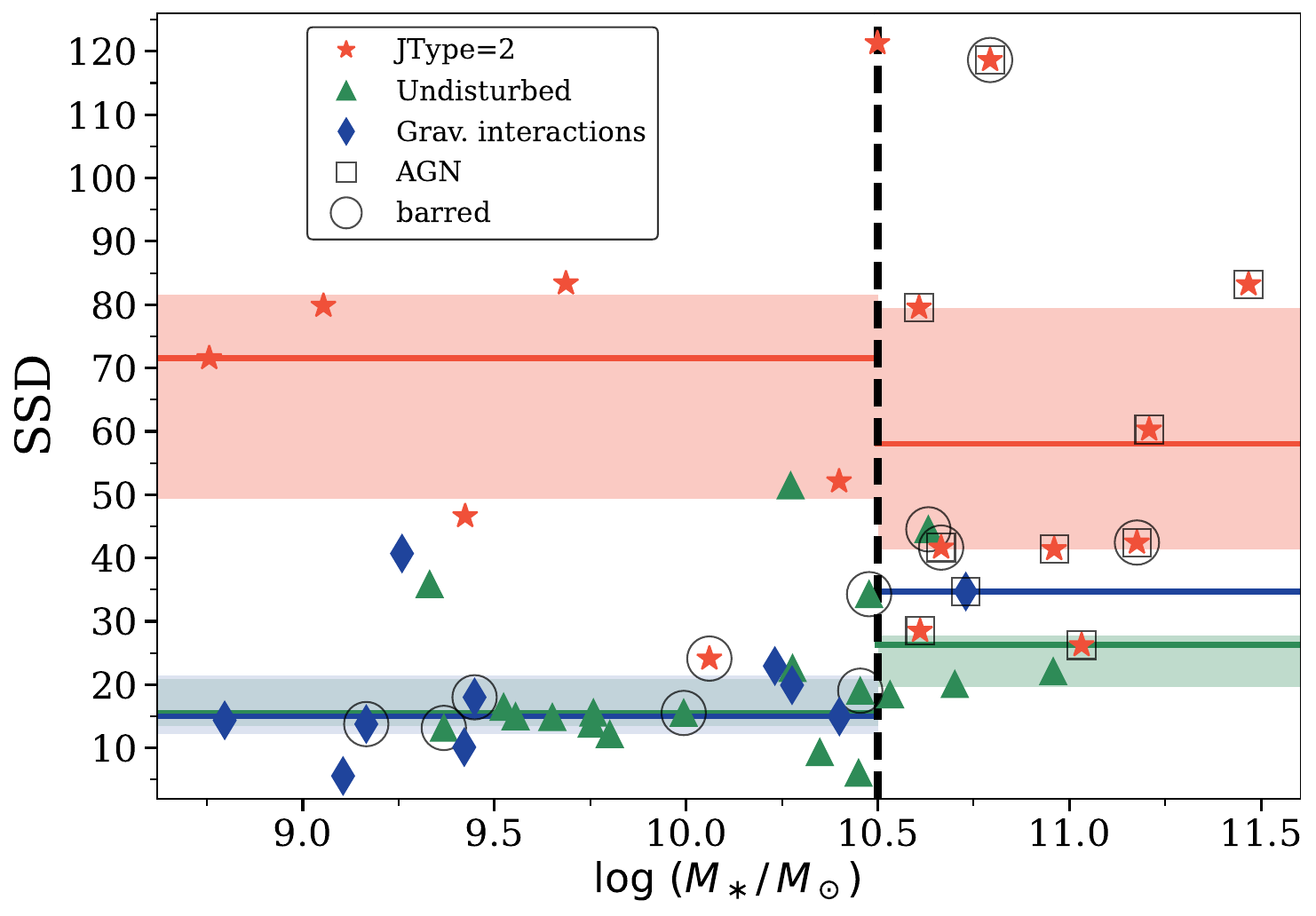}
    \vspace{-0.5cm}
    \caption{SSD values as a function of galaxy stellar mass. Galaxy types are represented by different symbols: JType~=~2 (red stars), undisturbed (green triangles) and gravitational interactions (blue diamond). If a galaxy hosts an AGN and/or exhibits a bar, an additional square or circle is overplotted, respectively. The vertical black dashed line marks the division between mass regimes at $\log (M_{\star} / M_{\odot}) = 10.5$. Solid lines and shaded areas indicate the median and interquartile range of SSD values for each galaxy type, respectively, computed within their respective mass regime.}
    \label{fig:cfr_mass}
\end{figure}

Figure \ref{fig:cfr_mass} shows the distribution of SSD values for undisturbed, JType~=~2 and gravitationally interacting galaxies,
now rearranged according to their stellar masses. The figure is divided into two panels: the left-hand panel displays the SSD distribution for galaxies in the lower-mass regime ($M_{\star} < 10^{10.5}\,M_{\odot}$), while the right-hand panel shows those in the higher-mass regime ($M_{\star} > 10^{10.5}\,M_{\odot}$).
As in Fig.~\ref{fig:moneyplot}, the solid lines and shaded regions indicate the median and interquartile range of SSD values for each galaxy type, respectively.

In the right panel, JType~=~2 galaxies remain segregated from both undisturbed and gravitationally interacting galaxies.
The median SSD values and their interquartile values in the high-mass regime, along with the number of galaxies per type, are:
JType~=~2 (9 galaxies): 58 (41, 80); gravitational interactions (1 galaxy): 35; Undisturbed (4 galaxies): 26 (20, 28).
Similarly, for the low-mass regime we have:
JType~=~2 (7 galaxies): 72 (49, 82); Gravitational interactions (9 galaxies): 15 (12, 21); Undisturbed (15 galaxies): 16 (13, 21).

Fig.~\ref{fig:cfr_mass} shows the SSD measure remains effective at distinguishing JType~=~2 galaxies from both gravitationally interacting and undisturbed galaxies.
For example, if we define a quantity $\Delta \mathrm{SSD} \equiv \mathrm{SSD}_{\mathrm{Jtype=2}} - \mathrm{SSD}_{\mathrm{Grav.}}$, this value is $\Delta \mathrm{SSD} = $\,40 when all galaxies are considered, with $\Delta \mathrm{SSD} = $\,57 in the low-mass regime and $\Delta \mathrm{SSD} = $\,32 in the high-mass regime, although for this case there is only gravitationally interacting galaxy and therefore we can not draw statistically robust arguments from this comparison. 

While the separation between gravitationally interacting and RPS galaxies appears less pronounced at higher stellar masses ($M_{\star} > 10^{10.5}\,M_{\odot}$), the median SSD values in the lower mass bin still differ by a factor of 1.2.
This suggests that the distinction observed in Fig.~\ref{fig:moneyplot} and Fig.~\ref{fig:cfr_types} is not driven by galaxy properties, but rather reflects the dominant physical mechanism affecting each galaxy. Future tests of the mass dependence, or lack thereof, with larger samples will be very useful in determining the statistical significance of this relation in mass-matched samples.
 
Additionally, in general, no clear trend is observed between SSD and the presence of either AGN or bars, even though we do not have enough statistics to perform a more in-depth analysis. For AGNs, we adopt the classification from \citet{Peluso2022}, which used the [\ion{N}{ii}]-based Baldwin–Phillips–Terlevich \citep[BPT;][]{Baldwin1981} diagnostic diagram to identify Seyfert-like radiation fields in GASP galaxies. For the bars, we use the classification by \citet{Sanchez-Garcia2023}, which combines visual inspection with detection of anomalous PA and eccentricity profiles in isophotes fitted to $I$-band images, to determine the presence or absence of stellar bars.

\section{Variations in the initial settings on SSD measurements}
\label{sec:testing_method}
In this section, we discuss the impact of modifying the number of slices used to measure the Lagrangian radii, as well as the effect of adjusting the SFRD fractions that define our selected set of Lagrangian radii. Additionally, we also introduce an objective metric applied to our sample to identify the fiducial bin width adopted throughout this work.

We measure the SSD values of undisturbed and JType~=~2 galaxies for eight different bin widths (5$^{\circ}$, 10$^{\circ}$, 15$^{\circ}$, 20$^{\circ}$, 30$^{\circ}$, 45$^{\circ}$, 60$^{\circ}$ and 90$^{\circ}$). For each configuration $j$, the two samples were compared adopting a Monte Carlo approach. We draw $N=5000$ realizations of the SSD measurements of JType~=~2 (group $A_j$) and undisturbed (group $B_j$) galaxies from a normal distribution centered on the observed value with a scale given by their SSD uncertainties. The Hodges–Lehmann (HL) estimator \citep{Hodges1963} was then computed as:

\begin{equation}
    \mathrm{HL}(A, B) = \frac{1}{N} \sum_{k=1}^{N} \mathrm{median}(A_k - B_k),
    \label{eq:WHL_estimator}
\end{equation}

\noindent with the index $k$ corresponding to each Monte Carlo realization. To quantify the separation between undisturbed and JType~=~2 galaxies, we define the following score:

\begin{equation}
    \mathrm{score} \equiv \frac{\mathrm{HL} \,(A_j, B_j)}{\sqrt{\sigma (A_j)^2 + \sigma(B_j)^2}}
    \label{eq:score}
\end{equation}

\noindent where $\sigma$ denotes the sample standard deviation and $A_j$, $B_j$ are the SSD measurements of JType~=~2 and undisturbed galaxies for a given configuration, respectively. Higher scores correspond to configurations that maximize the group separation while minimizing their combined scatter. We select as fiducial the configuration providing the highest normalized score among the eight combinations tested, which corresponds to a bin width of 20{\textdegree}.

We show in Sects.~\ref{subsec:vary_degbins} and \ref{subsec:vary_lagr_set} that our main result -- i.e. that the SSD measure of JType~=~2 galaxies is significantly higher than that of undisturbed galaxies or those undergoing gravitational interactions -- is robust against variations of the bin width and choice of Lagrangian radii within reasonable bounds.

\subsection{Changing the number of angles}
\label{subsec:vary_degbins}

The choice of the number of slices must take into account that if the slices are too narrow they may contain too few pixels, making the measurement of Lagrangian radii more unstable against noise contamination. On the other hand, if the slices are too wide, they may fail to capture the true morphological shape of the distribution of young and intermediate-age stars.

\begin{figure}
    \centering
    \includegraphics[width=\columnwidth]{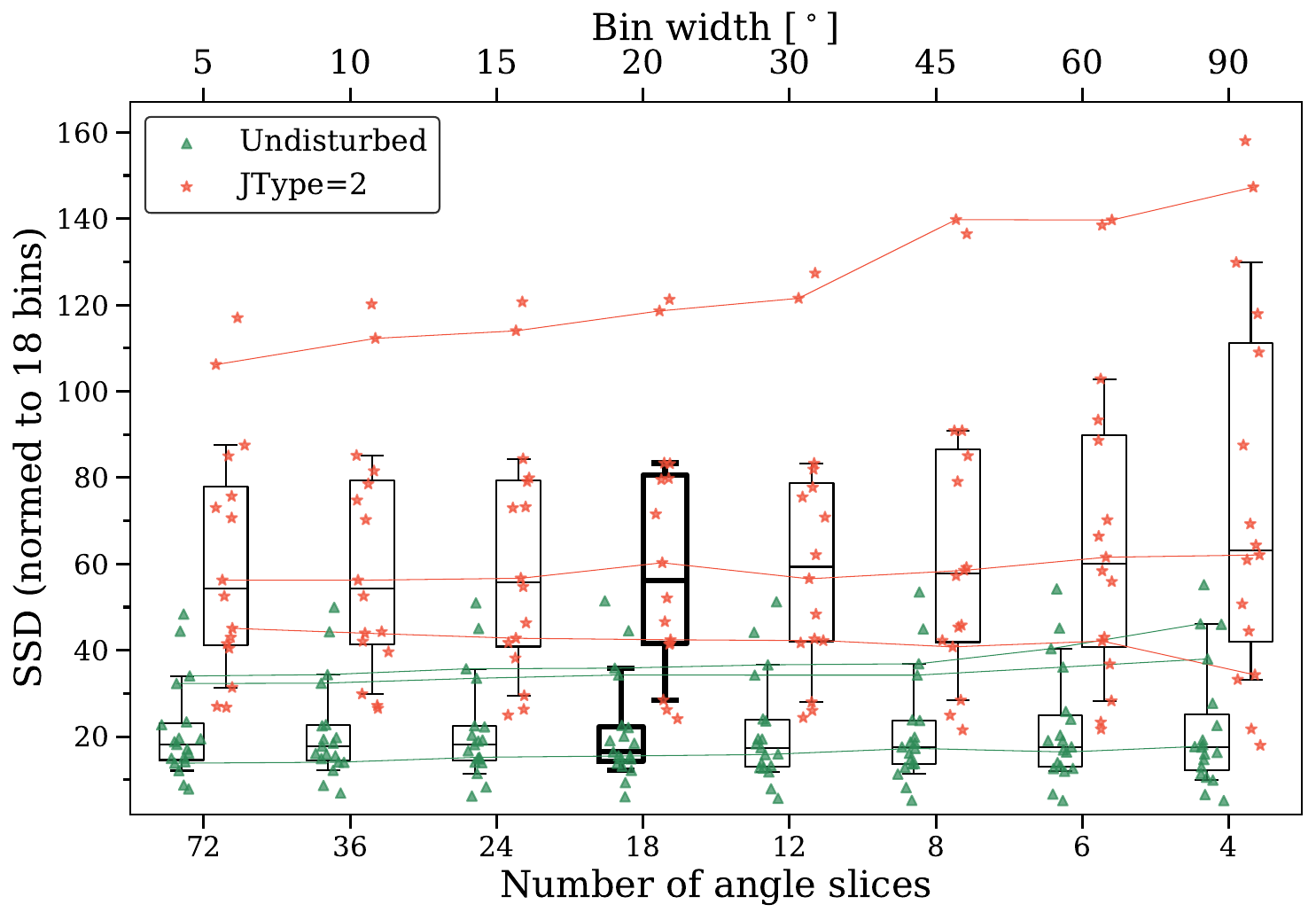}
    \caption{The SSD values of the undisturbed (green triangles) and JType~=~2 (red squares) galaxies, comparing the SSD measured using a different number (lower horizontal axis) or size (upper horizontal axis) of angle slices. Our fiducial choice of angle slices, shown with thicker lines, is 18 (20{\textdegree}). The measured SSD values are robust across a wide range of bin widths. For reference, we highlight the variation in SSD for three representative undisturbed galaxies (green solid lines) and three representative JType~=~2 galaxies (orange solid lines).}
    \label{fig:cfr_angles}
\end{figure}

In our fiducial SSD measure, we used 18 angle slices of 20{\textdegree} each in which we measured our Lagrangian radii. In Fig.~\ref{fig:cfr_angles}, we show that the SSD measure is quite robust to a large variation in the number of angle slices. We note that because the SSD is a sum of values calculated in each angle slice, for a fair comparison we normalize our SSDs to a fiducial 18 angle slices; for example, we divide our SSD using 72 angle slices by 4 and multiply our SSD value using 6 angle slices by 3. 

Remarkably, the median and interquartile ranges significantly overlap over the whole range of using 72 slices of 5\textdegree\ to 4 slices of 90\textdegree.
We stress that, although the difference between the median SSD values of the JType~=~2 galaxies is actually larger when only 4 angle slices are used in comparison to 18, as expected the scatter on the measured SSD values increases significantly compared to the cases using narrower angle steps, such that there is more overlap in the population distributions.
In addition to showing the overall distribution of SSD measures, in Fig.~\ref{fig:cfr_angles} we track the variation of individual randomly-selected galaxies across different bin widths. Most galaxies show very little variation in their SSD value, although a few JType~=~2 galaxies have their values increased with a decreasing number of bins.
We argue that this behavior is not surprising for galaxies with long tails, as the angle slice including the stripped tail drives up the SSD value because there the distribution of the youngest stellar populations extends farther than for the older populations.

\subsection{Changing the set of measured radii}
\label{subsec:vary_lagr_set}

\begin{figure}
    \centering
    \includegraphics[width=1\columnwidth]{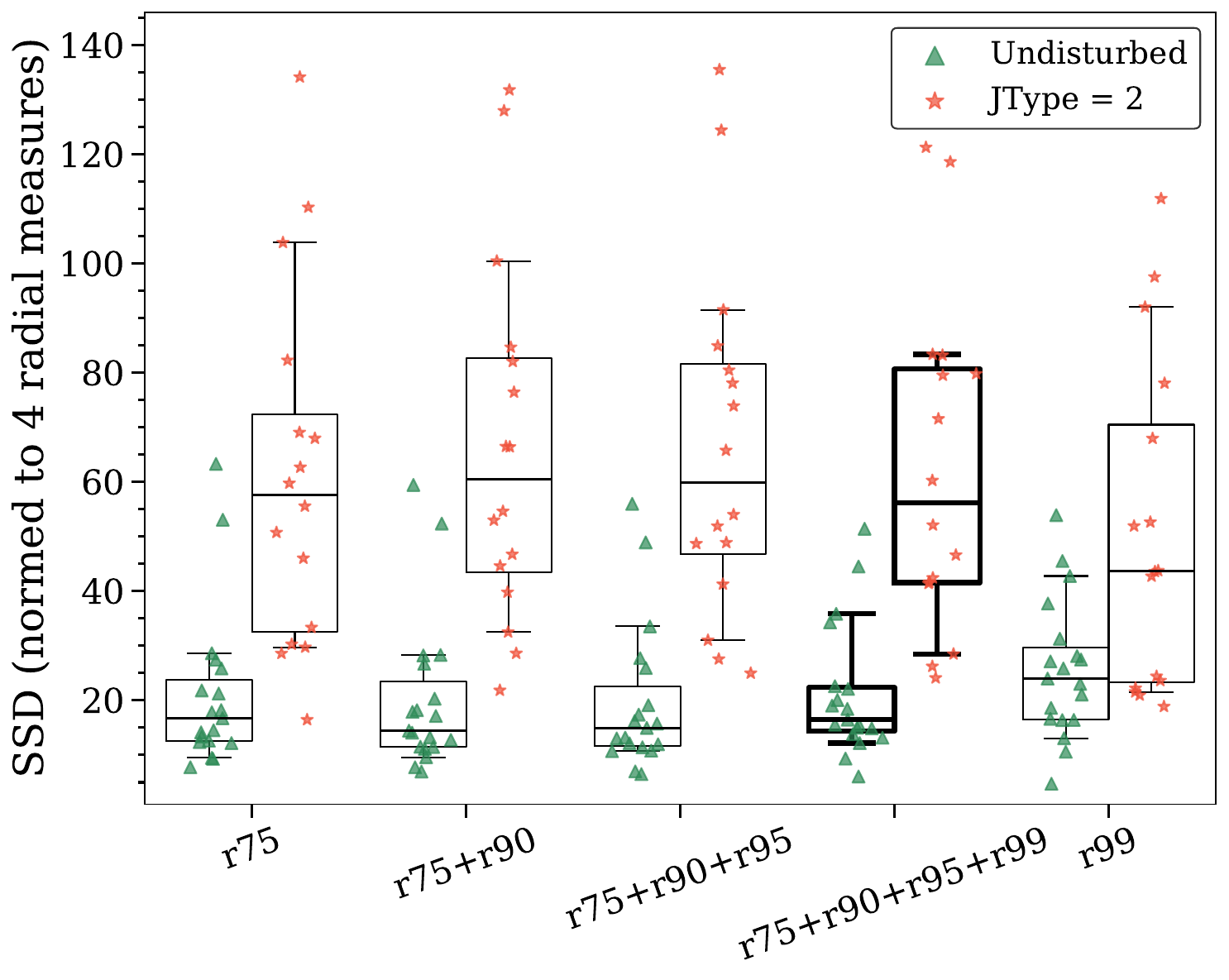}
    \vspace{-0.6cm}
    \caption{SSD values measured using different subsets of Lagrangian radii. Colors and symbols are as in Fig.~\ref{fig:cfr_angles}}
    \label{fig:cfr_radii}
\end{figure}

To maintain consistency with the methodology introduced in \citet{Smith2025}, we adopt the same set of Lagrangian radii ($r75+r90+r95+r99$). As discussed in Sect.~\ref{subsec:ssd_gasp}, we exclude $r50$ to avoid potential contamination from AGN or stellar bars, features that were not present in the simulated galaxies. In Fig.~\ref{fig:cfr_radii}, we test whether using different subsets of our Lagrangian radii affect our results. Again, we observe that the SSD robustly finds a difference between the undisturbed and JType~=~2 galaxy samples (no overlap in the interquartile ranges, although the 10\% -- 90\% ranges overlap), even when we use a single Lagrangian radius for the SSD\footnote{As with the angle bin comparison, the SSD values are normalized to using four Lagrangian radii. As a concrete example, the SSD measured using a single Lagrangian radius is multiplied by 4.}. As an example, Fig.~\ref{fig:cfr_radii} shows that $r75$ alone does a good job in separating undisturbed and JType~=~2 galaxies.

Interestingly, the $r75$ Lagrangian radius gives us the highest SSD median values and largest ranges for both the undisturbed and JType~=~2 galaxies. We hypothesize that this is because the central regions of galaxies tend to host older stars, so even when a galaxy has not been deeply affected by RPS, the $r75$ of the older stars should be smaller than that of the younger stars. We see this in about 75\% of the undisturbed sample.
While the youngest stars also tend to have a more extended $r99$, this difference is often smaller in the undisturbed sample, as seen in \galCONTROL in Fig.~\ref{fig:SSD_GASP}. In addition, once a galaxy is strongly stripped, the extended tail with newly-formed stars will increase all the Lagrangian radii in the young stellar population by construction.

\section{Application on imaging data}
\label{sec:single_band}
Determining whether a cluster galaxy is undergoing RPS often requires integral field unit (IFU) observations. IFU data enable a spatially resolved, spectroscopically based analysis of the host galaxy, providing detailed characterization of both its gas and stellar properties, as discussed in Sections \ref{sec:data_analysis} and \ref{sec:method}. 
Spatially resolved SFH allows for the identification of age distributions within the galaxy, which represent the key concept for the success of the SSD parameter in distinguishing strong RP effects from gravitational interactions. However, acquiring IFU data demands significant observational efforts, making it difficult to construct large spectroscopic samples of RPS galaxies.

Figures \ref{fig:SSD_GASP}, \ref{fig:moneyplot} and \ref{fig:cfr_types} express the ability of SSD to distinguish RPS and gravitational interactions in different GASP galaxies, serving as a test for the simulation-based idea proposed in S25, providing a quantitative assessment of the impact of RPS in different galaxies. However, this approach depends on the accuracy of full-spectral fitting techniques such as \sinopsis, which are essential to disentangle the contributions of different stellar populations to the observed continuum emission.

Given these considerations, this section aims to explore the applicability of SSD to single-band imaging observations, with the potential advantage of further application in large-scale imaging surveys, where SSD might serve as a selection tool to identify candidate RPS galaxies for spectroscopic follow-ups.
Considering the adaptations required to apply the method to imaging data -- which necessarily introduce additional parameters -- we do not test the sensitivity of the results to the input parameters, as was done in the case of the SFRD and presented in the previous sections. These aspects are deferred to future work. Here, we focus solely on assessing whether the method can distinguish undisturbed and gravitationally interacting galaxies from JType~=~2 galaxies when only imaging data is used.

\subsection{\texorpdfstring{H$\alpha$}{Hα} and broad-band imaging}
\label{subsec:ha_vs_bband}

We start by comparing narrow-band H$\alpha$ with the reddest broad-band filter fully covered by the MUSE wavelength range (Cousins-$I$). To replicate as closely as possible H$\alpha$ narrow-band imaging observations, we avoid using the emission-only and dust-corrected data cubes, as these were derived using spectroscopic techniques such as subtraction of the modeled stellar continuum and estimate of the color excess through the Balmer decrement, respectively \citep{Poggianti2017}. Instead, we use the data cubes corrected for Milky Way foreground extinction, produced using the dust maps from \citet{Schlafly2011} and assuming a CCM reddening law \citep{Cardelli1989}.
These data cubes are then used to generate broad-band and narrow-band H$\alpha$ images, which are created by integrating the flux by each passband transmission curve. To trace the distribution of the oldest stellar populations, we use the Cousins/$I$ passband, while for the case of the young stellar populations we adopt HST/WFC3 UVIS2 F680N filter. This passband covers rest-frame H$\alpha$ emission across the redshift range of $0.006 \lesssim z \lesssim 0.08$, thus including the majority of the galaxies analyzed in this work. The only exception is JO20 ($z = 0.14706$), which is therefore excluded from this analysis.
The procedure described in Sect.~\ref{subsec:ssd_gasp} to measure the SSD for each galaxy is then repeated using these single-band images, with the difference that, instead of deriving Lagrangian radii from the SFRD, we now use fluxes integrated over the corresponding filter passbands.

Unlike in the case of the SFRD maps, no prior S/N cut was applied, therefore, a masking step based on background subtraction is necessary in this case.
The background level in each image is estimated using \textsc{SExtractor} \citep{Bertin1996}, as implemented in the Python package \textsc{sep} \citep{Barbary2016}, adopting a square mesh of 32 pixels with a filter box of $8\times8$ pixels. All spaxels with fluxes below the global background level are masked. Narrow-band images were continuum-subtracted, with the continuum level estimated as the mean flux calculated from the Johnson/$B$ and Cousins/$I$ bands.
Uncertainties are computed following the procedure described in Sect.~\ref{subsec:ssd_gasp}, based on the flux errors from the datacube variance, which are propagated through the integration step.

\begin{figure}
    \centering
    \includegraphics[width=\columnwidth]{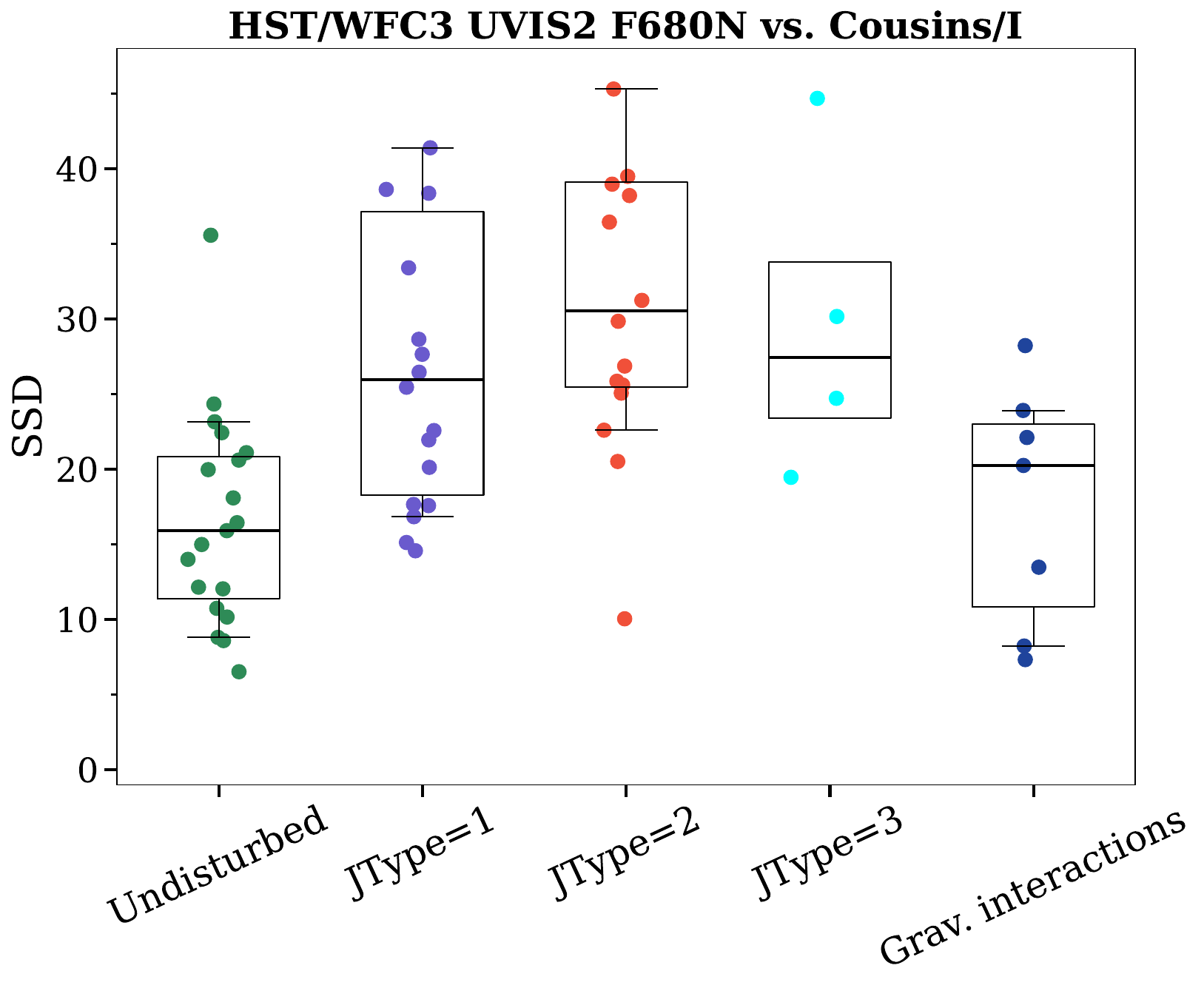}
    \caption{Distribution of SSD values for the galaxy sample analyzed in this work, now measured on MUSE-generated narrow-band H$\alpha$ and broad-band Cousins/$I$ images.}
    \label{fig:SSD_single_bands}  
\end{figure}

Figure~\ref{fig:SSD_single_bands} shows the distribution of SSD values, similarly to Fig.~\ref{fig:cfr_types}, but now measured from the images described above. As in Fig.~\ref{fig:cfr_types}, the median SSD value is higher for JType~=~2 galaxies, although the scatter of the derived values is evidently larger compared to the SFRD case, particularly for JType~=~1 and JType~=~2 galaxies. The median SSD values for each sample, along with their 25\%--75\% interquartile ranges, are as follows: Undisturbed: 21 (14, 29); JType~=~1: 22 (15, 33); JType~=~2: 43 (59); JType~=~3: 28 (22, 33); Gravitationally interacting: 14 (12, 22).
For comparison, the separation between the median SSD values of JType~=~2 and undisturbed galaxies is 40 in the SFRD case, while it decreases to 22 in the present case.
This reduction is an expected outcome of our deliberate choice not to correct the H$\alpha$ emission-line fluxes for underlying stellar absorption or ISM dust attenuation, in order to reproduce as closely as possible the conditions of H$\alpha$ narrow-band observations. Even so, the difference in SSD between JType~=~2 galaxies and the other groups remains sufficiently large to distinguish them.

\subsection{Blue and red broad-band imaging}
\label{subsec:blue_vs_red_bbands}
\begin{figure}
    \centering
    \includegraphics[width=\columnwidth]{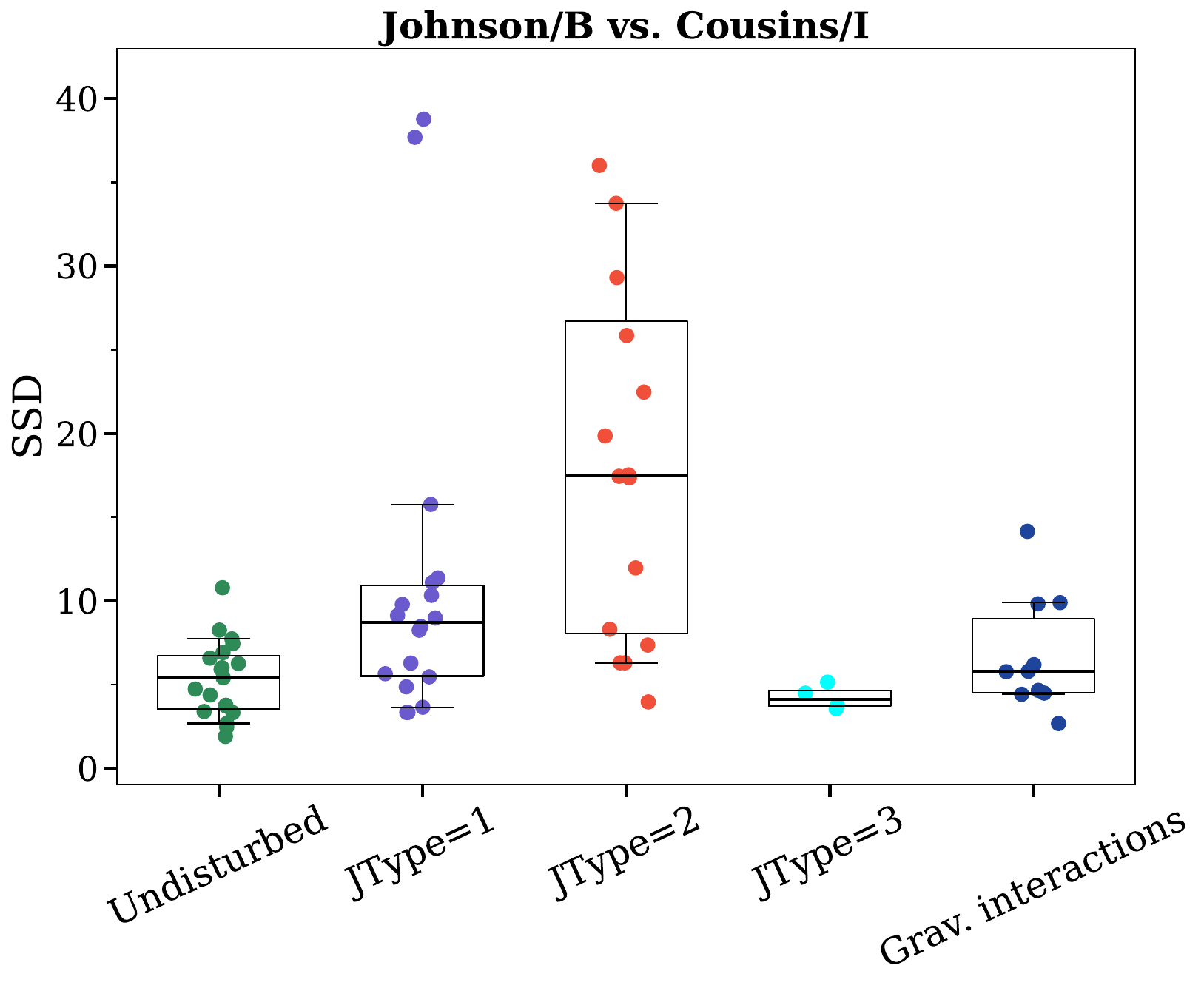}
    \caption{Same as Fig.~\ref{fig:SSD_single_bands}, but now SSD was measured using MUSE-generated broad-band images.}
    \label{fig:SSD_single_bands2}  
\end{figure}

Although Fig.~\ref{fig:SSD_single_bands} demonstrates that the SSD measure remains effective in distinguishing strong RPS from tidal interactions using only imaging data, the requirement for narrow-band H$\alpha$ coverage still imposes an observational limitation, as it restricts the redshift range over which the SSD method can be applied in large-scale surveys. With this in mind, we perform an additional test using the Johnson-$B$ band as a tracer of the young stellar population in each galaxy, keeping the Cousins-$I$ band as the tracer of older populations.
The results are presented in Fig.~\ref{fig:SSD_single_bands2}.
Similarly to the previous case, JType~=~2 galaxies exhibit higher median SSD values than the other galaxy classes: Undisturbed: 7 (5, 8); JType~=~1: 9 (6, 13); JType~=~2: 20 (12, 25); JType~=~3: 7 (6,7); Gravitationally interacting: 7 (5, 10). Consequently, the separation between JType~=~2 and undisturbed galaxies, although existent, is smaller here than in the previous examples. While the SFRD-derived SSD values show a separation of 40, this difference decreases to only 12 in the present case. Although a separation is still discernible, all galaxies are now clustered toward lower SSD values, with a non-negligible overlap between JType~=~2 and gravitationally interacting systems, for instance.

The decreased sensitivity of the two broad-band imaging data to the spatial distribution of different stellar populations is not surprising, as both young and intermediate-age populations emit across the optical wavelength range. The dilution of distinct morphological features may simply result from the spatial resolution limitations of the MUSE instrument at the typical distances probed. Such differences might be more easily identified with deeper and higher-resolution observations. When computed across a broad passband, the stripped tails of RPS galaxies become fainter and less pronounced, making the features that allow the SSD technique to distinguish between tidal interactions and RPS much more subtle. For comparison with Fig.~\ref{fig:SSD_GASP}, we present the Lagrangian radii of \galCONTROL and \galRPS in Appendix~\ref{app:c} for both cases.

It should be noted that the Johnson-$B$ passband used to trace the young stellar populations is only partially covered by the MUSE spectrograph wavelength range. Considering the redshift range of the analyzed sample, as well as the MUSE spectral coverage, the Johnson-$B$ passband captures rest-frame emission from $\sim$4480\,\AA\ to $\sim$5080\,\AA. With access to even bluer bands -- sensitive to the UV emission of younger stellar populations -- this difference would likely increase. Furthermore, we stress that the relatively small field of view of MUSE ($\sim$1\,arcmin$^{2}$) occasionally does not cover the full extent of the stripped tails, in which case the corresponding SSD would be underestimated. This limitation is naturally mitigated in large-scale surveys that observe substantially larger regions of the sky. Accounting for all these limitations is beyond the scope of the current analysis, and would certainly benefit from larger samples.

We highlight that a direct quantitative comparison between these values and those calculated using the SFRD can be misleading. First, the S/N threshold used in the SFRD calculation is not directly comparable to the flux threshold used in the single-band images, therefore the ``edges'' of each image will differ. Second, the use of H$\alpha$ narrow-band filters inevitably includes flux from the adjacent [\ion{N}{ii}] doublet, thereby leading to an underestimation of the intrinsic H$\alpha$ flux that may weaken its direct connection to the youngest stellar populations.
We present these examples of measuring the SSD using imaging data as a proof of concept that this analysis can be performed on a larger imaging dataset, thereby providing a potential tool to select RPS candidates for spectroscopic follow-up in a more systematic way than visual identification.

\section{Conclusions}
\label{sec:conclusion}
In a recent study, \citet[][S25 throughout this work]{Smith2025} introduced the Size-Shape Difference (SSD) measure to distinguish ram pressure stripping (RPS) from gravitational interaction as the dominant mechanism shaping the disk morphology. Using two simulated galaxies, each undergoing one of these mechanisms, they applied the SSD and showed that it successfully separates one case from the other. To do so, the SSD measure explores the distinct imprints left by each mechanism on the stellar age gradients of the host galaxy. Since gravitational interactions affect all stellar populations similarly, the resulting stellar age gradients tend to be symmetrically distributed. Strong RPS, on the other hand, truncates the gas disk on the leading edge and creates tails of cold gas on the trailing edge. As a result, younger stellar populations, related to the stars formed after RPS reshaped the disk morphology of the host galaxy, are distributed asymmetrically. 

In this work, we carried out an observational test for the SSD measure by applying it to 67 galaxies from the GAs Stripping Phenomena in Galaxies \citep[GASP;][]{Poggianti2017, Poggianti2025} sample. Below, we summarize the main findings of this paper.

\begin{enumerate}
    \item The SSD measure can convincingly distinguish morphological features produced by RPS in observed galaxies undergoing the peak of the stripping phase (JType~=~2 galaxies) from those produced in gravitationally interacting galaxies. This finding aligns well with the simulations shown in S25. The median SSD values, along with their corresponding interquartile ranges, are: JType~=~2 galaxies: 56 (42, 81); Gravitationally interacting galaxies: 16 (14, 22); Undisturbed galaxies: 16 (14, 22).
    
    \item JType~=~1 galaxies, those with weaker visual signature of RPS relative to the JType~=~2, exhibit  SSD values comparable to those of gravitationally interacting and undisturbed galaxies. In the case of the truncated disks (JType~=~3), the SSD values are closer to those measured for JType~=~2 galaxies. This is expected, as disk truncation is a morphological transformation leaving imprints on the SFRD across different epochs.

    \item We investigate the impact of changing the age interval that traces the older stellar populations. As a test, we used the age interval of $570 < t\,[\mathrm{Myr}] < 6 \times 10^{3}$.
    While individual SSD values can shift, the mean distribution of SSD values remains very similar, with the interquartile ranges of the SSD values within each of most galaxy categories showing an overlap.
\end{enumerate}

These results are promising and support the application of the SSD parameter to large-scale imaging surveys, at least for identification of the most striking cases of RPS. 
We calculated the SSD using alternative tracers for the young and older stellar populations in two representative galaxies, to provide a proof of concept for whether single-band imaging can also distinguish RPS from gravitationally interacting galaxies. 
When using H$\alpha$ narrow-band and Cousins-$I$ broad-band imaging as tracers of the young and old stellar populations, respectively, the SSD still distinguishes between the two environmental mechanisms, although the gap between the two systems is reduced.

Although this gap is reduced when only broad-band imaging data is used, we emphasize that this test relied on Johnson-$B$ broad-band imaging as a tracer of the young stellar population. Further testing should be conducted using near-ultraviolet (NUV) imaging instead, particularly when higher spatial resolution is available. Should a significant difference then emerge, this technique could be automated and potentially serve as a valuable tool to facilitate spectroscopic follow-up for integral field unit (IFU) observations of RPS galaxies with bright and extended optical tails.

\section*{Data availability}
The full version of Table~\ref{tab:minitab} is available in electronic form at the 
CDS via anonymous ftp to 
\href{https://cdsarc.u-strasbg.fr}{\texttt{cdsarc.u-strasbg.fr}} (130.79.128.5) or via 
\href{https://cdsarc.u-strasbg.fr/cgi-bin/qcat?J/A+A/}{\url{https://cdsarc.u-strasbg.fr/cgi-bin/qcat?J/A+A/}}.

\begin{acknowledgements}
We thank the referee for the constructive comments that helped improve this paper.
A.E.L., R.S., B.V. and P.C.C. thank the Center for Computational Astrophysics of the Simons Foundation for a very pleasant and productive stay. A.E.L. and B.V. acknowledge support from the INAF GO grant 2023 ``Identifying ram pressure induced unwinding arms in cluster spirals'' (P.I. Vulcani).
A.E.L. and B.V. acknowledge the CINECA (Italy) for awarding time in the LEONARDO supercomputer under the ISCRA initiative, providing high‐performance computing resources and support that were used in this work.
A.E.L. thanks R.S. de Souza for the insightful discussions on statistical methods.
R.S. and Y.L.J. acknowledge support from the Agencia Nacional de
Investigación y Desarrollo (ANID) through Basal project FB210003, FONDECYT Regular projects 1241426 and 123044, and Millennium Science Initiative Program NCN2024\_112.
This project has received funding from the European research Council (ERC) under the European Union's Horizon 2020 research and innovation programme (grant agreement No. 833824).
J.F. acknowledges financial support from the UNAM-DGAPA-PAPIIT IN110723 grant, Mexico.
L.M. acknowledges support from the Croatian Science Foundation, project HRZZ-MOBDOK-2023-8006.
\end{acknowledgements}

\vspace{-0.8cm}
\bibliographystyle{aa}
\bibliography{references}

\begin{appendix}
\section{Inclination of the disk}
\label{app:inclination}

The disk inclination with respect to the observer's line of sight is an additional factor potentially affecting the SSD measurement. For instance, in S25 it was shown that if a model disk suffering RP is strongly inclined, its SSD value approaches that of an isolated control galaxy seen face-on.

In all previous comparisons of SSD values across different galaxy types, disk inclinations were not taken into account. While \citet{Franchetto2020} estimated the approximate inclination of stellar disks in the GASP sample, the ionized tail in jellyfish galaxies is not necessarily aligned with their stellar disks, and observational determination of its inclination with respect to the line of sight is challenging. In this section, we examine the effects of applying inclination corrections to the measured SSD values, adopting the simplifying assumption that the tail always aligns with the stellar disk.

To correct for inclination ($i$), we use the estimated values reported in \citet{Franchetto2020}, including position angles (PAs). Galaxies lacking estimates for either PA or inclination are excluded from the corrections and comparisons presented in this section. The correction procedure begins with a rotation in the $x-y$ plane by the galaxy PA, after which a stretching factor of $1 / \cos(i)$ is applied purely along the projected $x$ direction of each Lagrangian radius.
We stress that although this method accounts for the geometric inclination of the stellar disk, in highly inclined systems the SFHs are derived from spectra that have been intrinsically azimuthally averaged when projected into 2D. This averaging makes it more difficult for \textsc{sinopsis} -- or any other full spectral fitting technique -- to accurately disentangle the mixed stellar populations, as briefly mentioned in Sect.~\ref{subsec:ssd_comparison}.
Finally, the success of the SSD method also depends on the orientation of the ICM–galaxy interaction. In particular, if stripping occurs along the observer line of sight, the spatial distributions of the younger and older stellar populations become azimuthally blended. As a result, distinguishing between stars formed from the stripped material and those within the stellar disk becomes difficult.

\begin{figure}
    \centering
    \includegraphics[width=\columnwidth]{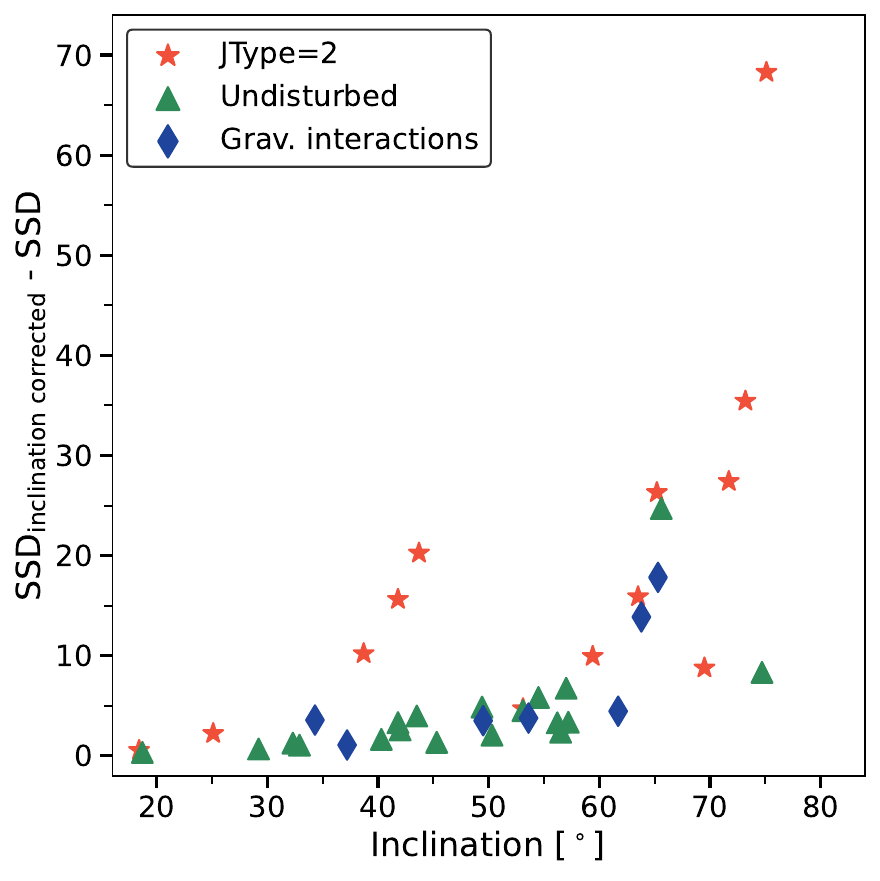}
    \caption{Effect of applying an inclination correction to the SSD measure for JType~=~2 (red stars), undisturbed (green triangles) and gravitationally interacting galaxies (blue diamonds) as a function of stellar disk inclination \citep[as reported by][]{Franchetto2020}.}
    \label{fig:cfr_incl}
\end{figure}

Figure \ref{fig:cfr_incl} compares SSD values with and without an inclination correction. The $y$-axis shows the difference in SSD measurements after applying the correction, plotted as a function of galaxy inclination. Similar to the findings in S25, the difference remains small ($\lesssim 20$) for undisturbed and gravitationally interacting galaxies, whereas for many JType~=~2 galaxies the difference is significant, reaching $\gtrsim 40$ in a few cases. Remarkably, the difference increases with disk inclination. The trend observed in Fig.~\ref{fig:cfr_incl} suggests that when the inclination of the stellar disk is accounted for -- despite the challenge of determining the inclination of the ionized tail observationally -- the clustering of jellyfish galaxies towards higher SSD values in comparison to the tidal interacting case in the diagram shown in Fig.~\ref{fig:moneyplot} can become even more pronounced. Therefore we consider our fiducial choice to not include an inclination correction to be conservative.

\section{\texorpdfstring{Using H$\alpha$-derived SFRDs to trace the youngest stellar population}{}}
\label{app:Appendix_Ha}

To model the stellar emission, \sinopsis measures the equivalent widths (EWs) of both emission and absorption features in selected continuum-dominated wavelength regions, searching for the combination of Simple Stellar Populations (SSPs) that provides the best fit to the observed spectrum. In the particular case of the youngest age bin ($t \leqslant 20$\,Myr), OB stars are still present and emit a sufficient number of energetic photons ($\lambda_{\gamma} \leqslant 912\,$\AA) to ionize H, leading to the emergence of several nebular emission lines in the observed spectrum.
The intensity of these lines, especially H$\alpha$, can be used to constrain the youngest stellar populations. \sinopsis incorporates this by computing predicted line intensities calculated from the Spectral Energy Distribution (SED) of $t \leqslant 20$\,Myr SSP models through \textsc{cloudy} \citep{Ferland2013} photoionization models \citep{Fritz2017}. Extinction is estimated from the observed H$\alpha$/H$\beta$ emission line ratio, assuming an intrinsic value of 2.863 \citep{Osterbrock2006}, and is applied to the models following the selective extinction hypothesis \citep{Calzetti1994}.

Although one would expect both SFR estimates to be consistent, it is important to note that, by construction, \sinopsis associates H$\alpha$ emission exclusively with ionization from young and massive stars. This assumption can lead to an overestimation of the derived SFRs in some regions -- including the stripped tail -- if, for example, the observed recombination lines are instead powered by other ionizing mechanisms, such as Low-Ionization Nuclear Emission-line Regions (LINERs). Additionally, the extinction estimate relies on the observed H$\alpha$/H$\beta$ line ratio. However, H$\beta$ is intrinsically fainter than H$\alpha$ -- even more when dust attenuation is present. In the outermost regions of the stripped tails, H$\beta$ is often undetected. Although \sinopsis imposes an upper limit on the allowed extinction, the detection of H$\alpha$ without H$\beta$ can mimic highly attenuated emission.
In light of these considerations, this section is dedicated to testing the effects of using SFRs derived directly from H$\alpha$ emission to trace the spatial distribution of the youngest stellar populations in GASP galaxies.

\begin{figure}[!ht]
    \centering
    \includegraphics[width=\columnwidth] {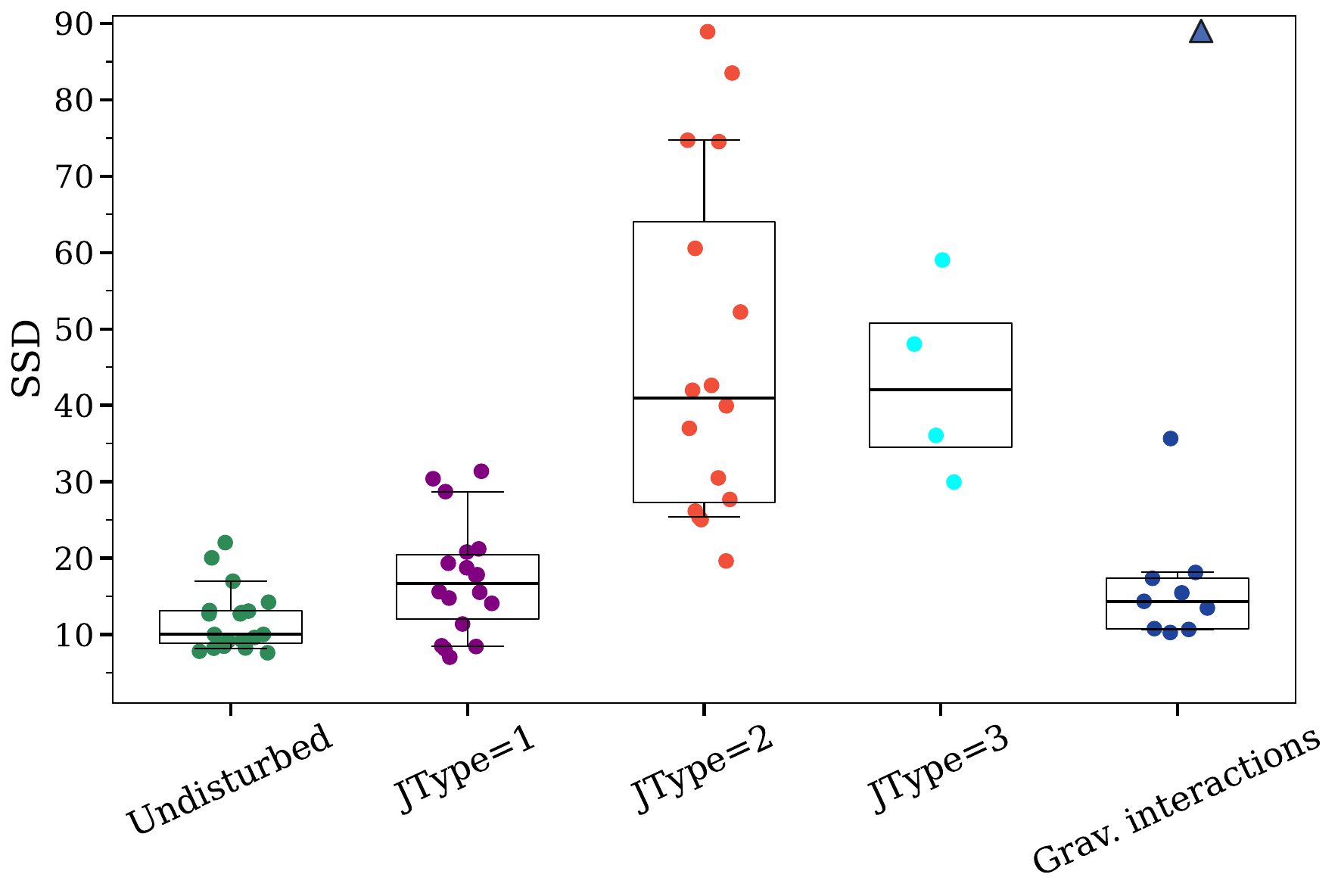}
    \caption{Same as Fig.~\ref{fig:cfr_types}, but with SSD values computed using SFRs derived directly from dust-corrected H$\alpha$ emission-line fluxes. The triangle marks JO190, which has an out-of-scale SSD value. The galaxy is still included in the calculation of the median for the gravitational interacting sample.}
    \label{fig:wBPT_cfr_types}
\end{figure}

Emission-line fluxes and associated uncertainties are estimated using the \textsc{kubeviz} code \citep{Fossati2016}, after applying an average-filter to the data cubes along the spatial direction with a 5$\times$5 kernel and subtracting the stellar continuum using \textsc{pPXF}. For the latter, \textsc{miles} stellar templates \citep{Vazdekis2010} were adopted, with the \verb|PADOVA 2000| isochrones \citep{Girardi2000}.
The SFRs are calculated using the calibration coefficient from \citet{Wilkins2019}, which are based on the \textsc{bpass} \citep{Eldridge2017} binary stellar evolution models and assume a \citet{Chabrier2003} stellar IMF:

\begin{equation}
    \log\bigg[\frac{\mathrm{SFR}}{M_{\odot}\,\mathrm{yr}^{-1}} \bigg] = \log\bigg[\frac{L(\mathrm{H}\alpha)}{\mathrm{erg}\,\mathrm{s}^{-1}} \bigg] - 41.37
    \label{eq:SFR_ha}
\end{equation}

\noindent where the calibration above assumes an upper stellar mass limit of $M_{\mathrm{upp}} = 100\,M_{\odot}$ and Solar metallicity. The emission line fluxes used to calculate $L(\mathrm{H}\alpha)$ were corrected by dust attenuation, as detailed in \citet{Poggianti2017}.

To assess how the SSD values of the 67 galaxies analyzed in this work are affected when using SFRs derived from Eq.~\ref{eq:SFR_ha}, we reproduce Fig.~\ref{fig:cfr_types} with the SFRD$_{\mathrm{H}\alpha}$. The results are shown in Fig.~\ref{fig:wBPT_cfr_types}. Only spaxels with (S/N)$_{\mathrm{H}\alpha} \geqslant 3$ are considered, and spaxels classified either as Seyfert-like or LINER-like in the BPT-[\ion{N}{ii}] diagnostic diagram \citep{Baldwin1981} are excluded\footnote{The only exception is JW100, for which SF spaxels are identified using the BPT-[\ion{S}{ii}] diagram instead, as the diagnostic diagram based on the [\ion{N}{ii}]~$\lambda$6583/H$\alpha$ ratio is  contaminated by a residual sky line \citep[see][for more details]{Poggianti2019a, Poggianti2019b}.}.
We also tested the results without applying the BPT diagnostic diagrams to identify SF spaxels, obtaining nearly identical SSD values. This is consistent with the findings of \citet{Poggianti2019a}, who reported that, for the majority of galaxies analyzed, at least 75\% of the H$\alpha$ emission in the stripped tails is due to star formation.

Comparing the results with those shown in Fig.~\ref{fig:cfr_types}, we find that the undisturbed sample remains nearly unchanged in individual SSD values, and consequently, in its median SSD. The JType~=~1 category shows a slight increase in the median SSD, though this variation
is within the uncertainties and consistent with previous values.
Nonetheless, the median SSD for the gravitationally interacting sample remains well below that of the JType~=~2 galaxies. The latter show a systematic decrease in their median SSD from 56 to 41, although this variation lies within the uncertainty.

Interestingly, the median SSD for the JType~=~3 category shifts from $\sim$27 in Fig.~\ref{fig:cfr_types} to $\sim$42.
While two galaxies in this category exhibit nearly identical values before and after the change (JO23: 32.3/29.9; JW108: 62.0/59.0), the two others show significant increases (JO10: 18.0/48.0; JO36: 21.5/36.1).
Unfortunately, the small sample size prevents drawing firm conclusions about the median behavior of truncated disk galaxies when replacing SFR1 with SFRs derived directly from H$\alpha$.
Overall, despite individual variations in SSD values, the SSD method remains effective at distinguishing galaxies undergoing strong RPS from those experiencing gravitational interactions. Regardless of the method used to derive the SFRs, truncated disks remain exhibiting higher SSD values compared to the undisturbed systems, showing that the SSD method is also sensitive to past RPS events.

\section{Illustrative example of Lagrangian radii behavior measured on imaging data}
\label{app:c}

\begin{figure}[!h]
    \centering
    \includegraphics[width=\columnwidth]{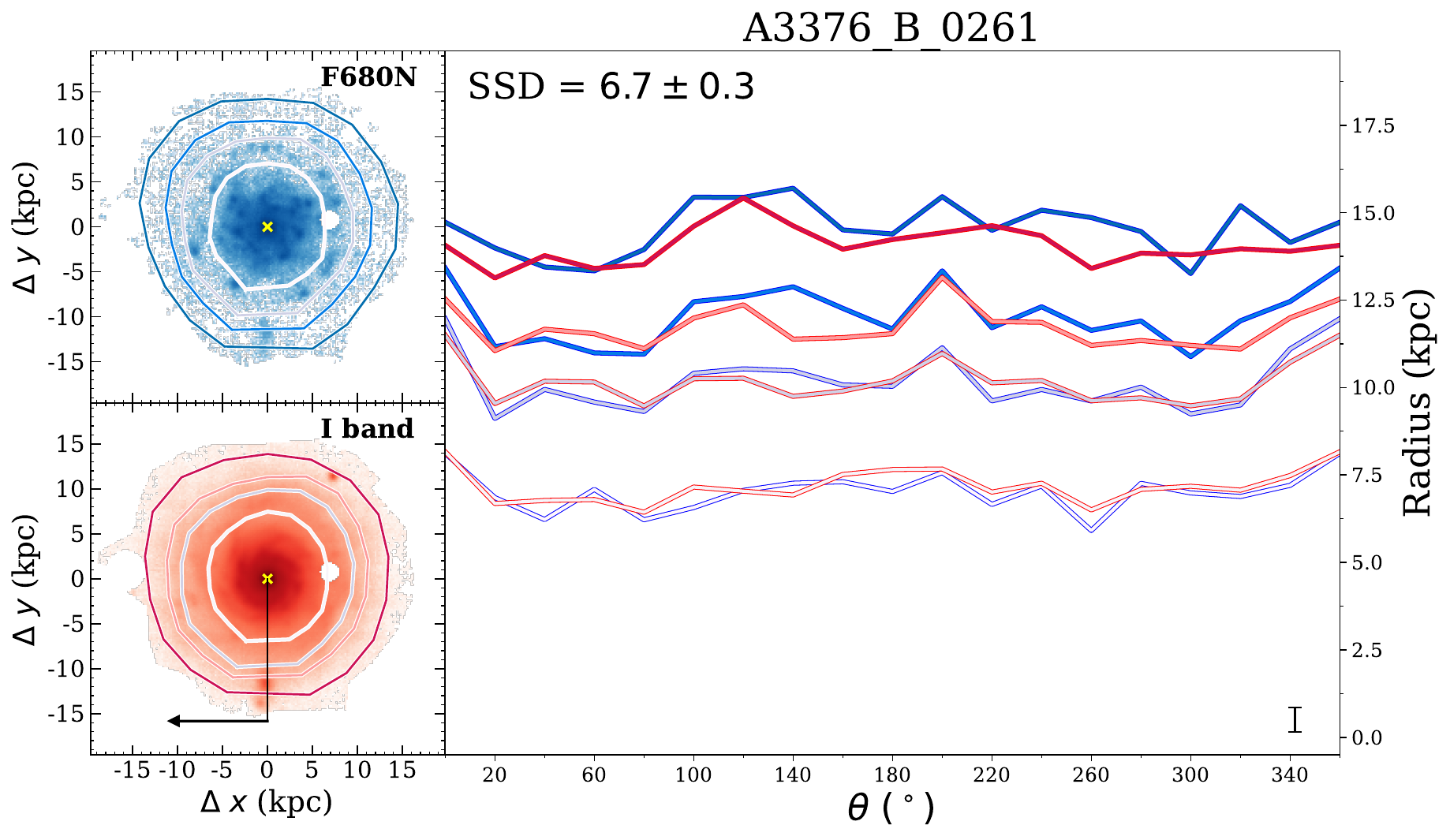}
    \vspace{0.3cm}
    \includegraphics[width=\columnwidth]{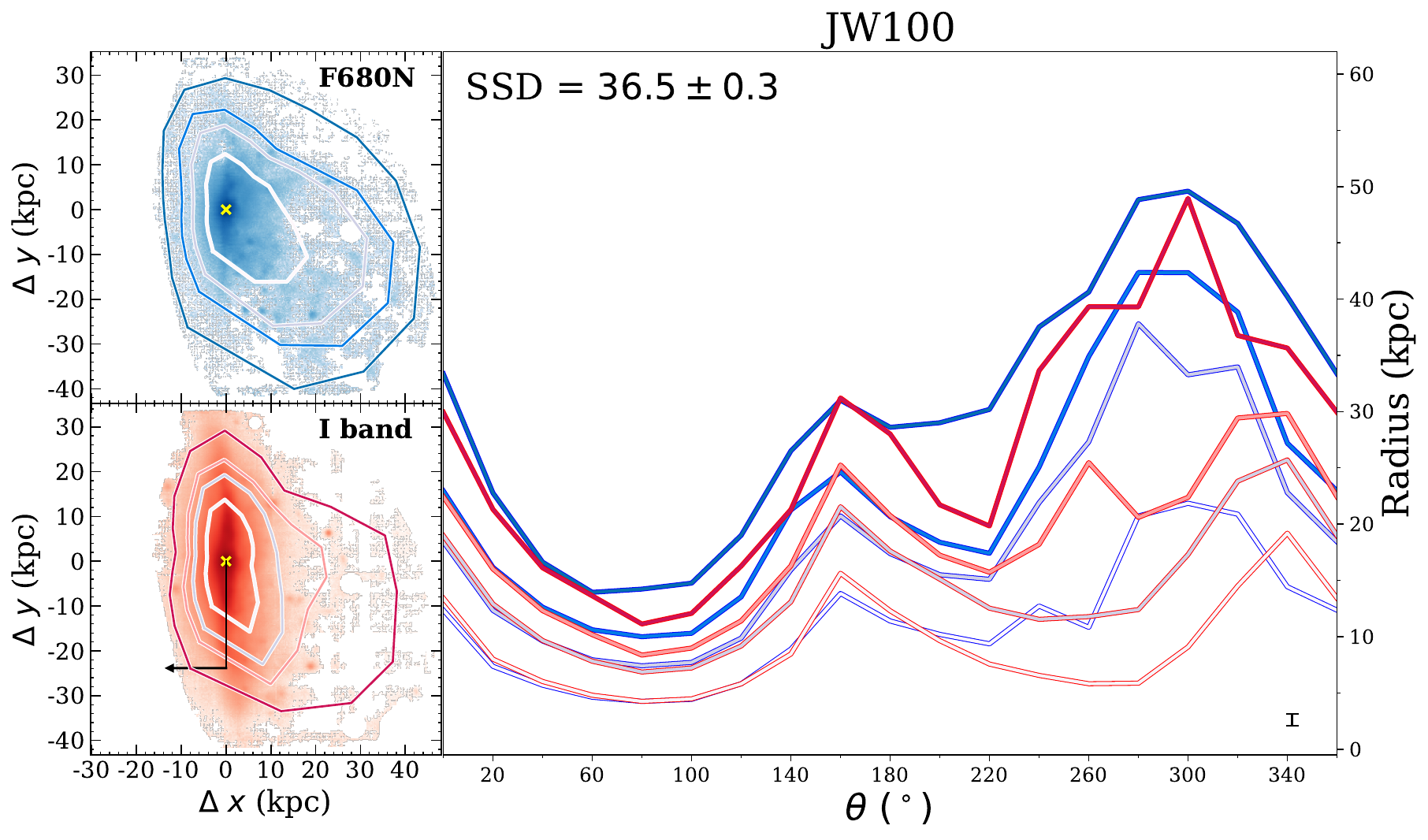}
    \vspace{-0.8cm}
    \caption{Same as Fig.~\ref{fig:SSD_GASP}, but using HST/WFC3 UVIS2 F680N to trace the young stellar populations and Cousins-$I$ to trace the intermediate-age stellar populations in each galaxy.}
    \label{fig:lgr_nb_bb}
\end{figure}

\begin{figure}
    \centering
    \includegraphics[width=\columnwidth]{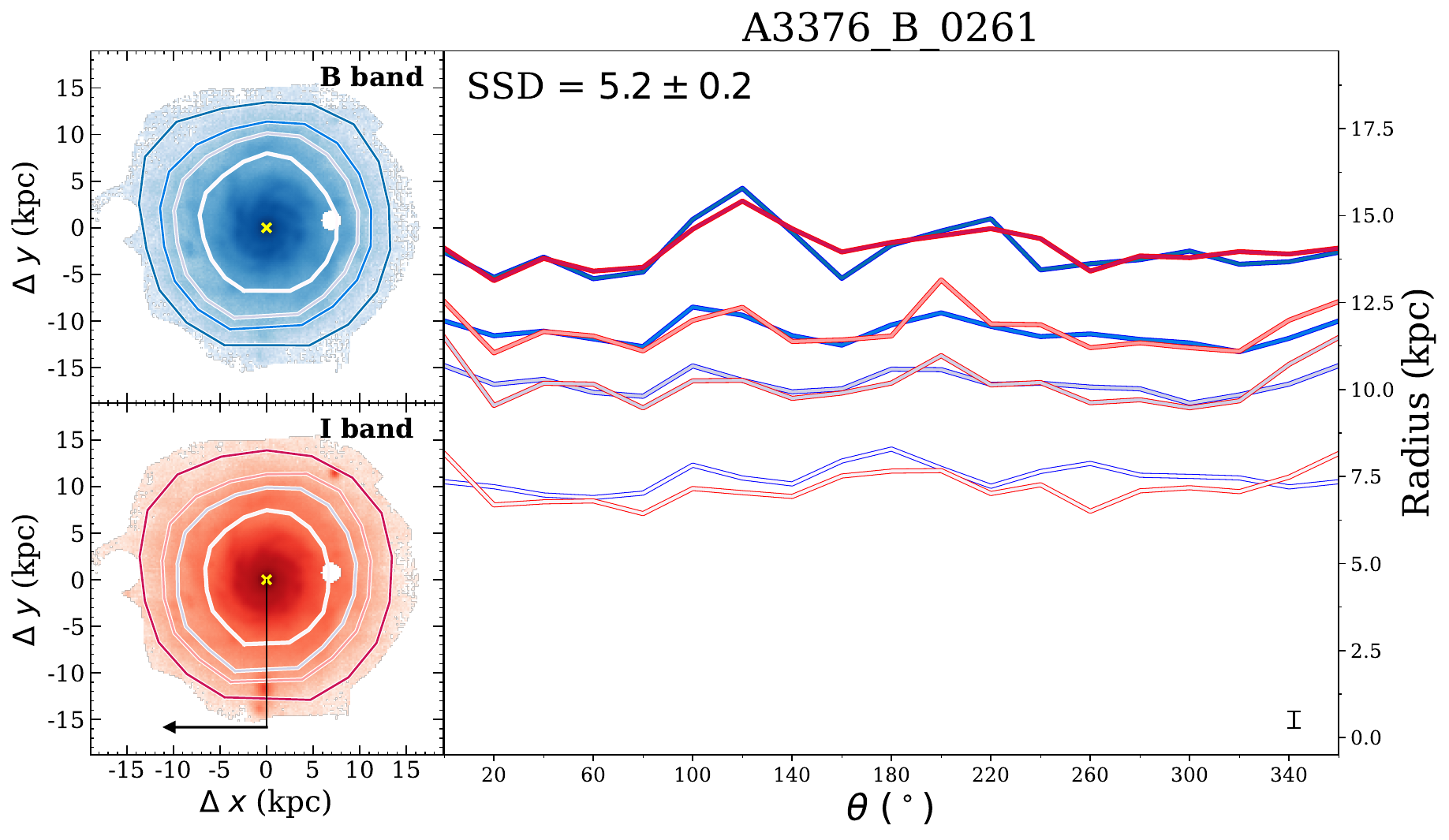}
    \vspace{0.3cm}
    \includegraphics[width=\columnwidth]{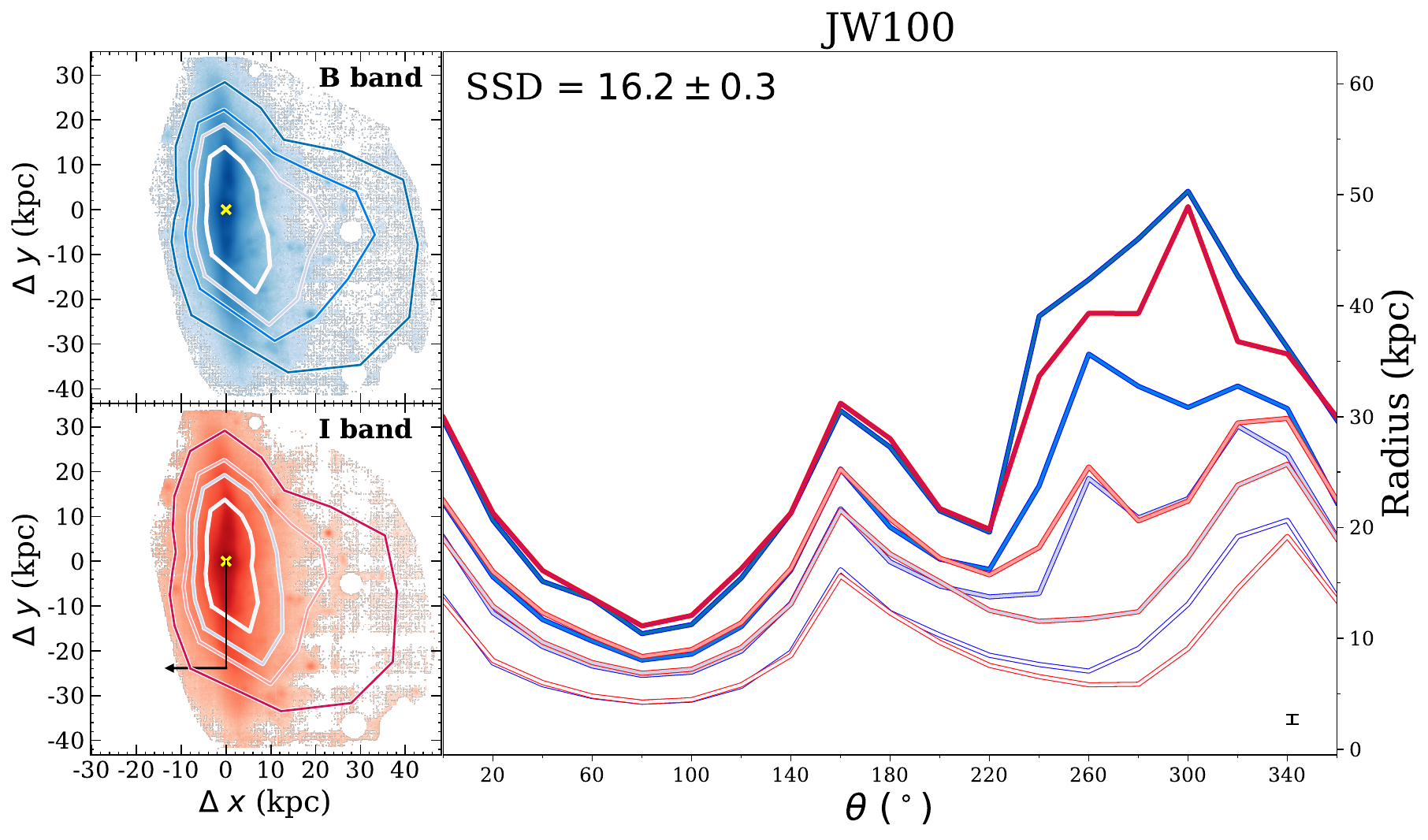}
    \vspace{-0.8cm}
    \caption{Same as Fig.~\ref{fig:SSD_GASP}, but using Johnson-$B$ broad-band image and Cousins-$I$ to trace the young and intermediate-age stellar populations in each galaxy, respectively.}
    \label{fig:lgr_bb_bb}
\end{figure}

We present the Lagrangian radii overlaid on the corresponding images in Figs.~\ref{fig:lgr_nb_bb} and \ref{fig:lgr_bb_bb}. Unlike in the SFRD case, the truncation at the leading edge of the disk is no longer apparent (in practice the Lagrangian radii at these position side of the \galRPS became nearly identical). The stripped tail, on its turn, is hardly visible in the B band and narrow-band H$\alpha$, but it differs from the I band image mostly at $\theta \sim 220^{\circ}$, slightly raising the measured SSD value.

\end{appendix}
\end{document}